%% file: main.tex
\documentclass[journal,dvipsnames]{new-aiaa}

\usepackage[utf8]{inputenc}
\usepackage{graphicx}
\usepackage{amsmath}
\usepackage{tabularx}
\usepackage{siunitx}
\usepackage{longtable,tabularx}
\usepackage{caption}
\usepackage{float}
\usepackage[colorinlistoftodos]{todonotes}
\usepackage{comment}
\usepackage{subcaption}
\usepackage{array,ragged2e}
\usepackage{color, colortbl}
\usepackage{tikz}

\usepackage{amssymb}
\usepackage{orcidlink}

\input{functions/markers}

\title{Analysis of Circulation Control Jet Bi-Stability on a Wing Section at Transonic Speeds via Dynamic Mode Decomposition}

\author{\orcidlink{0000-0001-6508-2084} Dor Polonsky \footnote{Rafael, Aerodynamics Group, Haifa, Israel, dmpwork0@gmail.com} }
\newcommand{\orcid}[1]{\href{https://orcid.org/0000-0001-6508-2084}{\textcolor[HTML]{A6CE39}{\aiOrcid}}}

\affil{Rafael Advanced Defense Systems Ltd}

\begin{document}
\maketitle

\begin{abstract}
\input{sections/abstract}

\end{abstract}

\input{sections/nomenclature}

\section{Introduction}
\label{sec:introduction}
\input{sections/intorduction}

\section{Geometric Model}
\label{sec:geometry}
\input{sections/geometry}

\section{Numerical Verification \& Validation}
\label{sec:numericsNgrid}
\input{sections/numerics_VnV}

\section{Results \& Discussion}
\label{sec:results}
\input{sections/results}

\section{Conclusions}
\input{sections/conclusions}
\label{sec:conclusions}

\clearpage
\bibliography{references}

\end{document}

%% file: functions/markers.tex
\definecolor{navyblue}{rgb}{0.00,0.45,0.74}
\definecolor{dandelion}{rgb}{0.93,0.69,0.13} 
\definecolor{matgreen}{rgb}{0.39,0.83,0.07}

\DeclareRobustCommand{\bluecirc}{\tikz \draw[navyblue] (0,0) circle (.5ex);}
\DeclareRobustCommand{\stndcirc}{\tikz \draw[black] (0,0) circle (.5ex);}
\DeclareRobustCommand{\redcirc}{\tikz \draw[red] (0,0) circle (.5ex);}
\DeclareRobustCommand{\redcircfilled}{\tikz \draw[red,fill=red] (0,0) circle (.5ex);}
\DeclareRobustCommand{\dandicirc}{\tikz \draw[dandelion] (0,0) circle (.5ex);}

\DeclareRobustCommand\sampleline[1]{%
  \tikz\draw[#1] (0,0) (0,\the\dimexpr\fontdimen22\textfont2\relax)
  -- (2em,\the\dimexpr\fontdimen22\textfont2\relax);%
}

\DeclareRobustCommand\samplelineShort[1]{%
  \tikz\draw[#1] (0,0) (0,\the\dimexpr\fontdimen22\textfont2\relax)
  -- (0.75em,\the\dimexpr\fontdimen22\textfont2\relax);%
}

%% file: sections/abstract.tex
The phenomenon of stable lift oscillations occurring on an elliptic wing section utilizing circulation control at transonic speeds was evaluated using numerical simulations. As the momentum of the jet increases beyond a prescribed magnitude, periodic detachment occurs from the trailing-edge. This behavior conforms to a bi-stable state, consistent with prior experimental observations.
Analysis by both steady and unsteady Reynolds Averaged Navier-Stokes calculations showed that the effect is decoupled from the dominant upstream shockwave. This indicates that the jet can no longer augment the wing’s circulation, marking the termination of circulation control. Furthermore, the results confirm that the absence of the downstream separation bubble acts as the catalyst for this detachment. 
Dynamic Mode Decomposition analysis revealed that the bi-stability is driven by a pressure feedback between the trailing-edge shockwave and a downstream pressure bubble. A secondary feedback governs the pressure redistribution during the detachment cycle.
It was concluded that the pressure-dominant nature of the bi-stability allows it to be captured using relatively simple methods such as URANS, and even approximated through a Reduced Order Model comprising only 2\% of the total modes, encapsulating 25\% of the modal influence and reconstructing the pressure field with 98\% accuracy.

%% file: sections/nomenclature.tex
\section*{Nomenclature}
{\renewcommand\arraystretch{1.0}
\noindent\begin{longtable*}{@{}l @{\quad=\quad} l@{}}
$A$                                 & time propagation matrix \\
$\tilde{A}$                         & POD projected time propagation matrix \\
$b$                                 & DMD initial conditions vector ,(Pa)\\
$c$                                 & airfoil chord length, (m) \\
$c_d$                               & drag coefficient \\
$c_l$                               & lift coefficient \\
$C_p$                               & pressure coefficient \\
$\delta C_p$                        & pressure coefficient reconstruction error \\
$C_{\mu}$                           & jet momentum coefficient \\
$\Delta c_l$                        & incremental lift coefficient due to jet blowing \\
$f$                                 & harmonic frequency, (Hz) \\
$f_n$                               & natural frequency, (Hz) \\
$I$                                 & modal influence \\
$I_\mathrm{id}$                     & identity matrix \\
$\dot{m}$                           & mass flow rate, (kg/s) \\
$M$                                 & Mach number \\
$p$                                 & pressure, (Pa) \\
$Q_{dyn}$                           & dynamic pressure, (Pa) \\
$r$                                 & mode number \\
$r_e$                               & trailing-edge major axis, (m) \\
$Re_c$                              & chord-based Reynolds number \\
$R_{g}$                             & specific gas constant, (J/(kgK)) \\
$t$                                 & time ,(s) \\
$t_c$                               & bi-stability period, (s) \\
$t_{c,0}$                           & initial detachment time, (s) \\
$\Delta t$                          & time step,(s) \\
$T$                                 & temperature ,(K) \\
$V$                                 & right singular matrix \\
$W$                                 & POD projected eigenvector matrix \\
$y^+$                               & non-dimensional wall distance \\
$x$                                 & chordwise coordinate, (m) \\
$X$                                 & data matrix \\
$\alpha$                            & angle of attack, (deg) \\
$\gamma$                            & specific heat ratio \\
$\zeta$                             & modal damping ratio \\
$\kappa$                            & reduced frequency \\
$\lambda$                           & continuous eigenvalue \\
$\mu$                               & discrete eigenvalue \\
$\nu_j$                             & jet isentropic velocity, (m/s) \\
$\xi$                               & exponential decay\\
$\rho$                              & density ,(kg/m$^3$) \\
$\Sigma$                            & singular value matrix \\
$\tau$                              & normalized cycle time \\
$\Phi$                              & modal spatial distribution \\
$\Psi$                              & left singular matrix \\
\end{longtable*}}

\noindent $\, $ \textbf{Subscripts}
\setlength\LTleft{0cm}
\begin{longtable}{@{}l @{\quad=\quad} l@{}}
$0$                                                  & total  conditions\\
$p$                                                  & plenum conditions\\
$\infty$                                             & freestream conditions\\
$\mathrm{jet}$                                       & jet specific properties\\
\end{longtable}

%% file: sections/intorduction.tex
\lettrine{C}ontrol authority over aerial vehicles via active flow control (AFC) has garnered considerable attention, particularly through collaborations such as the 2010 DEMON~\cite{yarf-abbasiDesignIntegrationEclipse2007, crowtherIntegratedDesignFluidic2009} and the 2018 MAGMA demonstrators, conducted jointly by BAE Systems and the University of Manchester~\cite{warsopFluidicFlowControl2018, warsopNATOAVT239Task2019}. Similar efforts, including the DARPA-funded CRANE project~\cite{montoroDARPACRANEProgram2023}, further highlight the growing interest in AFC due to its potential to enhance flight capabilities, as demands for vehicles of high maneuverability, endurance and range are on the rise. Recent research continues to expand AFC capabilities~\cite{mcveighFullScaleFlightTests2011, linOverviewActiveFlow2016, keisarPlasmaActuatorApplication2019, polonskyNoiseReductionModel2023, polonskyBatteryPoweredHelicopterHover2023}, reinforcing its viability in meeting these demands.

Among AFC methods, Circulation Control (CC), which relies on the injection of a Coand\v a jet~\cite{coandaDeviceDeflectingStream}, holds promise for flight maneuvering. By injecting the jet onto a rounded trailing-edge, centrifugal pressure forces the jet to attach to the surface, entraining surrounding air. This shifts the rear stagnation point along the trailing-edge curvature, increasing circulation and enhancing lift. These fluidic effectors are simple and aerodynamically efficient, capable of achieving high lift coefficients. However, their performance is constrained by airfoil design limitations, particularly regarding trailing-edge radius and duct height~\cite{englar1971design}. Despite these constraints, the MAGMA demonstrator successfully utilized CC alongside Fluidic Thrust Vectoring as a primary maneuvering mechanism, even without strictly adhering to optimal design margins. This suggests that there remains significant potential for further exploration of CC design space.

For transonic speeds, CC implementation faces additional challenges. The presence of wave drag and potential shockwave buffets~\cite{giannelis2017review}, the requirement for higher jet velocities to achieve meaningful effects~\cite{loth1984circulation}, and the limited number of experimental studies compared to subsonic conditions~\cite{alexanderTrailingEdgeBlowing2005, milholenEnhancementsFASTMACCirculation2013, milholenHighReynoldsNumberCirculation2012, chanTransonicDragReduction2017} underscore the need for robust numerical validations. Research into transonic gust load alleviation~\cite{liAirfoilGustLoad2020a}, shape optimization~\cite{forsterMultipointOptimisationCoanda2016}, and CC aerodynamics for supercritical airfoils~\cite{forsterNumericalSimulationTransonic2015a, forster2017design, chenNumericalStudyLift2021} has demonstrated the potential of CC at these conditions.

Currently, transonic CC numerical research primarily relies on steady jet blowing. For a steady jet, its momentum, expressed as the momentum coefficient must be low enough to prevent premature jet separation, limiting its functionality. However, detachment behavior is not always consistent across different designs, particularly when using an elliptic trailing-edge, shown in early work by Englar~\cite{englarTwoDimensionalTransonicWind1970} to be favorable for transonic speeds. Experimental results by Alexander~\cite{alexanderTrailingEdgeBlowing2005} revealed that in some cases, jet separation is not immediate but is preceded by a lift plateau. Numerical validations of this behavior~\cite{forsterNumericalSimulationTransonic2015a, liAirfoilGustLoad2020a} indicate that attempting to maintain a steady jet within this plateau region leads to unsteadiness in both residuals and aerodynamic coefficients.

This study will show that the observed lift plateau is a consequence of periodic jet separation and reattachment. This bi-stability has been depicted previously in experiments regarding the use supersonic jets in quiescent air~\cite{jegede2016dual, robertson2017influence} and thus is an established phenomenon. Nevertheless, as verified throughout this paper, the current bi-stability is heavily influenced by the presence of the transonic free-stream and its consequent trailing-edge shockwave. Furthermore, this periodic behavior can be numerically captured by low-resolution methods such as URANS and analyzed using advanced data-driven methods based solely on its pressure field. Using methods such as Proper Orthogonal Decomposition (POD)~\cite{brunton2022data} enables the decomposition of the flow-field into its dominant coherent structures, while Dynamic Mode Decomposition (DMD)~\cite{kutz2016dynamic} projects these structures onto linearized dynamic modes, facilitating a deeper understanding of unsteady dynamics. These techniques have been extensively used to analyze periodic motion in transonic flows~\cite{santana2024analysis, poplingher2019modal, liu2018analysis}, allowing for the identification of key flow features and a more comprehensive study of the periodic detachment behavior.

DMD itself is derived by the eigen-decomposition of the system's dynamic operator. In practice however, the full decomposition is circumvented by projecting the operator unto the POD modes, obtained through Singular Value Decomposition (SVD), resulting in complex eigenvalues. Such values encode frequency, amplitude and decay rate information, in a similar manner to the classical Fourier transform. This technique effectively identifies low-rank structures within the data~\cite{brunton2022data}. Since DMD is purely data-driven, it does not require prior knowledge of system dynamics but relies on high-resolution, deterministic data, a requirement linked to its foundation in Koopman operator theory~\cite{koopman1931hamiltonian}. This constraint on time resolution can, however, be relaxed through sparsity-promoting methods such as compressive sensing~\cite{kutz2016compressed}. The applications of DMD to both analyze the system dynamics as well as to produce a Reduced Order Model (ROM), exemplifies the robustness of a general data-driven approach to system analysis. 

Since unsteady periodic motion can induce responses in aerodynamic forces as well as structural fatigue effects via aeroelastic coupling~\cite{dung2020research}, understanding the periodic detachment can impact transonic CC design methodologies. Thus, this study aims to validate the bi-stable phenomenon, elaborate on its underlying mechanisms, analyze whether an ROM can be used to approximate it, and assess its implications for circulation control. 

The paper is structured as follows: Section~\ref{sec:geometry} details the geometry of the model used for calculations, followed by the numerical verification and validation of both the steady and unsteady behavior in Section~\ref{sec:numericsNgrid}. The results in Section~\ref{sec:results} present the characterization of bi-stability using URANS, including its analysis, decomposition and reconstruction using DMD. Finally, the conclusions of the study are presented in Section~\ref{sec:conclusions}.

%% file: sections/geometry.tex
\begin{figure}[h]
    \centering
    \includegraphics[scale=0.7]{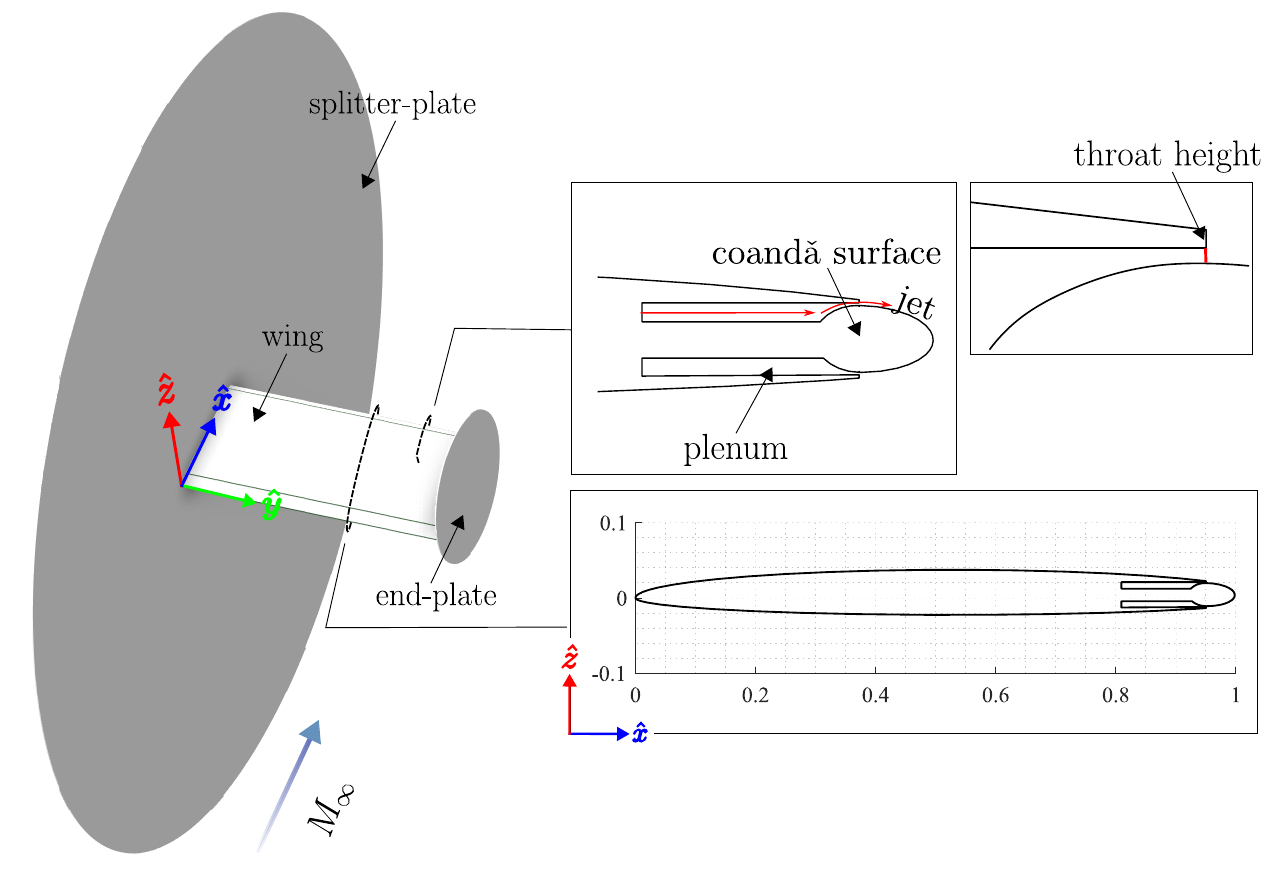}
    \caption{Geometry of the modeled TDT experimental wing.}
    \label{fig:Geometry}
\end{figure}

The wing cross-section used in this study, depicted in Fig.~\ref{fig:Geometry}, is based on the geometry employed in circulation control experiments conducted at NASA’s Transonic Dynamic Tunnel (TDT)~\cite{alexanderTrailingEdgeBlowing2005} and is employed as the validation case for the numerical calculations. The results obtained from the tunnel tests are used, and defined throughout the study as the "TDT tunnel data". The test wing featured a 6\% thick elliptic airfoil with $0.75\%$ camber and a nominal chord length of $c = 30$ inches which was used for computing aerodynamic coefficients.
The actual chord length, however, was $28.36$ inches due to a modification where the trailing-edge was cut at $0.9c$ and replaced with a dedicated Coand\v a surface. This replacement surface had an elliptic shape with a major-to-minor axis ratio of $2.98$, featuring a major axis radius $r_e$, of 1.37 inches.
The internal plenum, integrated within the wing, was designed to generate the jet required for circulation control. It featured a throat height of $0.012c$ and a length sufficient to produce a fully developed turbulent jet. 

The test wing spanned $60$ inches, corresponding to an aspect ratio of $2$, and was secured in place using a splitter-plate reinforced to the tunnel wall. To mitigate three-dimensional effects, an end-plate with a diameter of one chord in length was initially attached to the wingtip. However, numerical validations~\cite{liAirfoilGustLoad2020a, cruzAssessmentUnstructuredGridMethod2006, forsterNumericalSimulationTransonic2015a} conducted for transonic circulation control research, along with the author’s prior comparisons with preliminary calculations of the two-dimensional geometry, revealed that both the end-plate and splitter-plate significantly contributed to 3D effects. These effects were observed to influence the strength and position of the upper surface shock at the free-stream Mach number of 0.8. To account for these influences, both the end-plate and splitter-plate were explicitly modeled and their dimensions were selected based on previous numerical studies~\cite{forsterNumericalSimulationTransonic2015a}, with values of $1.1c$ and $6c$, respectively. The computational domain was represented as a cylindrical volume with a diameter of $10c$ extending $7c$ in the spanwise direction. To ensure accuracy, the numerical study was validated against the earlier stated experiment at the TDT, which was selected due to its comprehensive dataset covering a wide range of Mach numbers $M$, angles of attack $\alpha$, and jet momentum inputs.

%% file: sections/numerics_VnV.tex
\subsection{Computational Grid}
\label{subsec:grid}

\begin{figure}[H]
    \centering
    \includegraphics[scale=0.75]{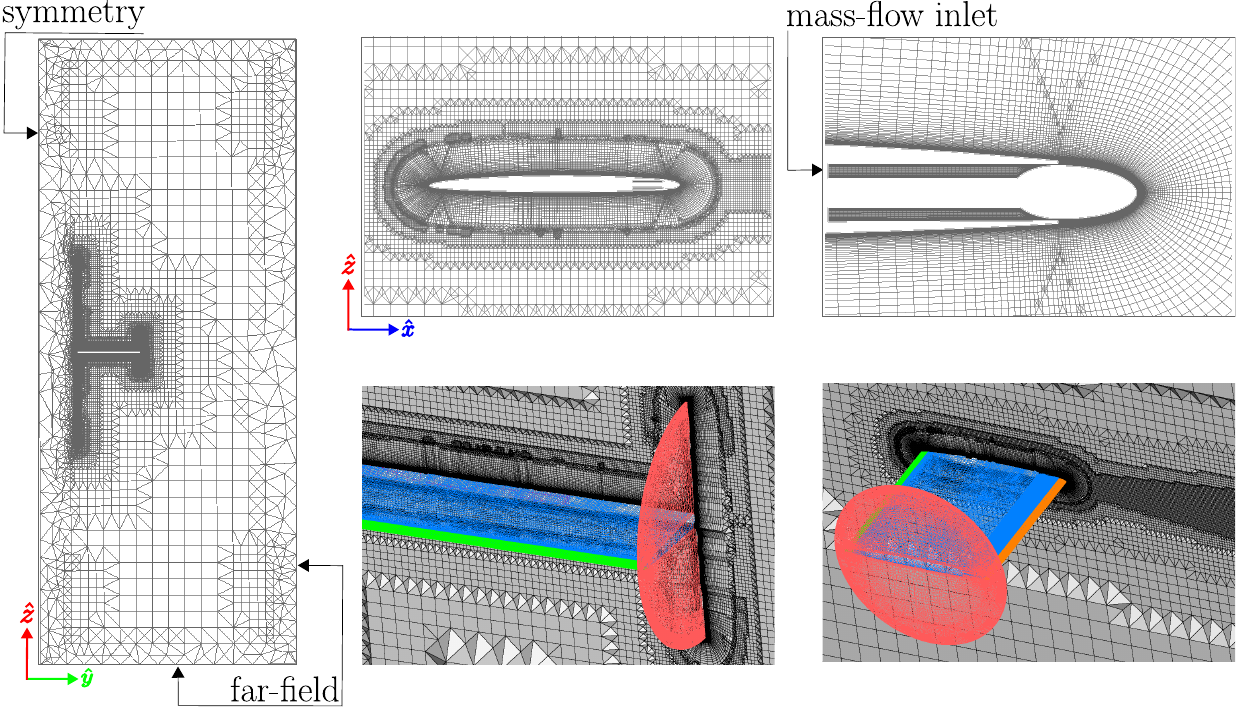}
    \caption{Computational domain illustrating the relevant boundary conditions and cross-sections of the grid.}
    \label{fig:Domain}
\end{figure}

\begin{figure}[H]
 \centering
 \begin{subfigure}[b]{0.49\textwidth}
     \centering
     \includegraphics[scale=0.6]{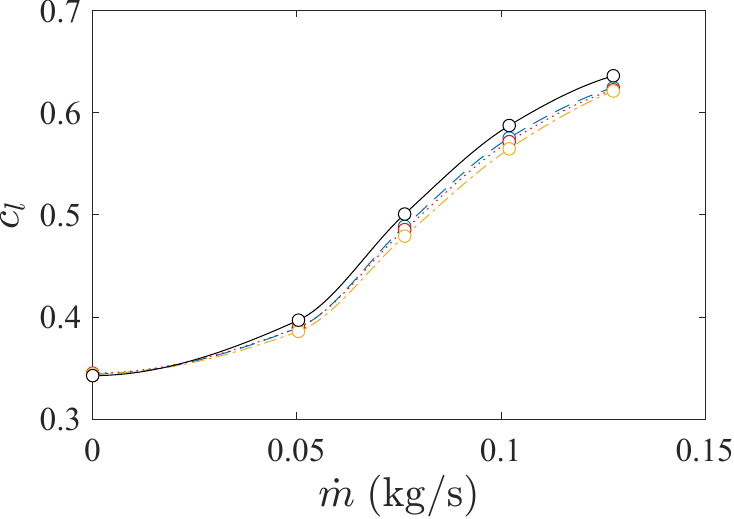}
     \caption{lift grid convergence}
     \label{}
 \end{subfigure}
 \hfill
 \begin{subfigure}[b]{0.49\textwidth}
     \centering
     \includegraphics[scale=0.6]{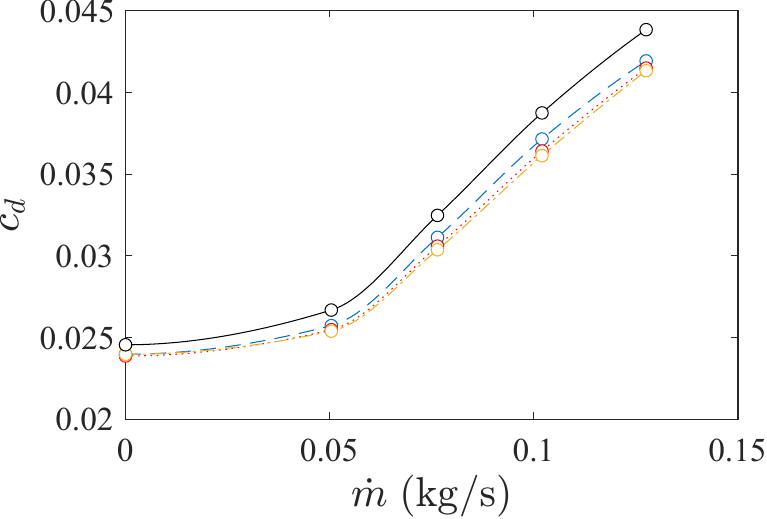}
     \caption{drag grid convergence}
     \label{}
 \end{subfigure}
    \caption{Convergence of the mid-span lift and drag coefficients with increasing grid count for different mass flow inputs at an angle of attack of 3$^\circ$ and a free-stream Mach number of 0.8. \\
    \fbox{\samplelineShort{}\stndcirc \samplelineShort{} $N = 1.3 \times 10^6$, \textcolor{navyblue}{\samplelineShort{dashed}\bluecirc \samplelineShort{dashed} $N = 1.9 \times 10^6$}, \textcolor{red}{\samplelineShort{dotted}\redcirc \samplelineShort{dotted} $N = 2.7 \times 10^6$}, \textcolor{dandelion}{\samplelineShort{dash pattern=on .5em off .2em on .05em off .2em}\dandicirc \samplelineShort{dash pattern=on .5em off .2em on .05em off .2em} $N = 4 \times 10^6$}}}
    \label{fig:GridVerification}
\end{figure}

The computational domain was constructed using an unstructured mesh generated with Fidelity$^\copyright$ Pointwise, employing the \textit{Voxel} algorithm. This resulted in a hybrid grid comprising hexahedral and tetrahedral cells, as illustrated in Fig.~\ref{fig:Domain}, which depicts some of the grid's cross-sections and associated boundary conditions. 
To study the effects of the grid while accurately capturing the underlying flow physics, an initial fine mesh consisting of $40 \times 10^6$ cells was utilized, and subsequently coarsened up to $13 \times 10^6$ cells. Across all grid configurations, the boundary layer was resolved using 80 layers growing at a rate of 1.1, while ensuring the first cell height corresponded to $y^+ \approx 1$. 
Four grid variants were developed, focusing on enhanced cell density near the wing, splitter-plate, end-plate, and near-wake regions. For each grid, the mid-span lift coefficient and drag coefficient were computed and compared as depicted in Fig.~\ref{fig:GridVerification}. Grid convergence was deemed achieved when the variation in these aerodynamic coefficients was within $\mathcal{O} \leq 1\%$. The final grid, consisting of $27 \times 10^6$ cells, was used for the remainder of the study.

\subsection{Numerical Setup}
\label{subsec:numerics}

The computational study was conducted using Ansys$^\copyright$ Fluent simulation software by solving the compressible Reynolds Averaged Navier-Stokes (RANS) equations.
Boundary conditions were set to replicate the experimental tunnel conditions~\cite{alexanderTrailingEdgeBlowing2005} with free-stream Mach number of $M_\infty=0.8$, static pressure $p_\infty = 13789.5 \; \mathrm{(Pa)}$ and temperature $T_\infty = 308.15 \; \mathrm{(K)}$ corresponding to a free-stream chord-based Reynolds number of $2.88\times 10^6$. The jet flow from the plenum was modeled as a mass-flow inlet boundary condition, determined by the required momentum coefficient $C_\mu = \dot{m}\nu_\mathrm{jet} / {Q_{\mathrm{dyn}} \, c}$, where $\dot{m}$ is the mass-flow to the plenum, $Q_\mathrm{dyn}$ is the free-stream dynamic pressure and $\nu_\mathrm{jet}$ is the isentropic jet velocity calculated by the expansion of the jet to the free-stream and defined as:
\begin{equation}
    \nu_\mathrm{jet} = \sqrt{2R_gT_{0,p} \left(\frac{\gamma}{\gamma -1}\right) \left(1 - \left(\frac{p_\infty}{p_{0,p}} \right)^{\frac{\gamma - 1}{\gamma}} \right)}
\end{equation}
where $R_\mathrm{gas}$ is the specific gas constant, $\gamma$ is the specific heat capacitance ratio and $T_{0,p}$, $p_{0,p}$ are the total temperature and total pressure at the plenum respectively. To maintain consistency with the experimental definition of the lift coefficient, $c_l$, the pressure was integrated along the mid-span of the wing:
\begin{equation}
    c_l = \frac{1}{c}\int_0^c (C_{p,\mathrm{low}} - C_{p,\mathrm{up}})dx 
\end{equation}
Where $C_{p,\mathrm{up}}$ and $C_{p,\mathrm{low}}$ are the upper and lower surface pressure coefficients respectively. 

For the numerical schemes, the inviscid fluxes were discretized using second order Roe flux difference splitting scheme and the solution gradients were computed using the Green-Gauss node based approach. Both the second order Spalart-Allmaras (SA) and $k-\omega$ SST turbulence models were tested to reduce uncertainty and their validation is expanded upon in the following section. 

\subsection{Turbulence Model \& Pressure Validation}

\begin{figure}[h]
     \centering
     \begin{subfigure}{\textwidth}
         \centering
         \begin{subfigure}[b]{0.49\textwidth}
             \centering
             \includegraphics[scale=0.6]{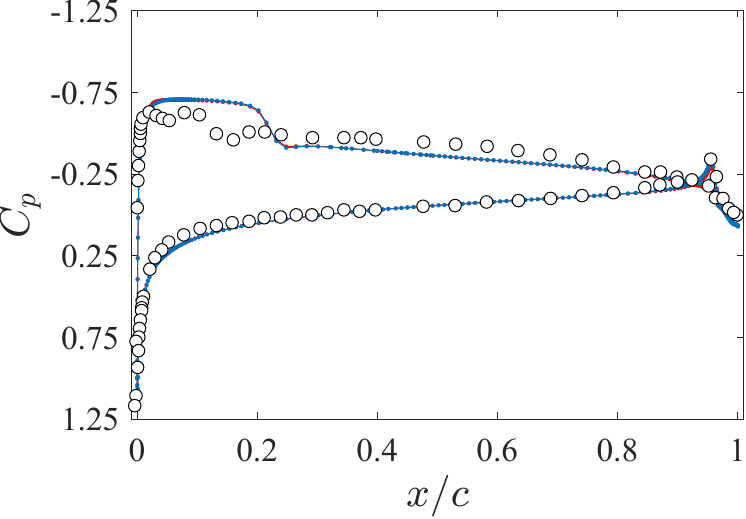}
         \end{subfigure}
         \hfill
         \raisebox{0.82cm}{
         \begin{subfigure}[b]{0.49\textwidth}
             \centering
             \includegraphics[scale=0.975]{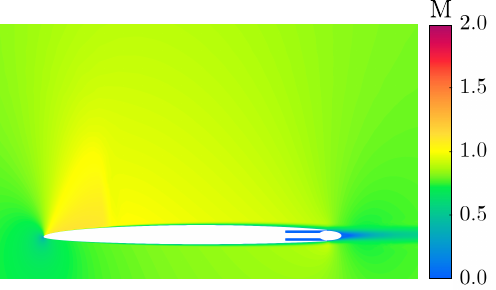}  
         \end{subfigure}}
         \caption{no jet blowing, $C_\mu = 0$}
         \label{fig:TurbValidationAOA3CMU0}
     \end{subfigure}
     \hfill
     \begin{subfigure}{\textwidth}
         \centering
         \begin{subfigure}[b]{0.49\textwidth}
             \centering
             \includegraphics[scale=0.6]{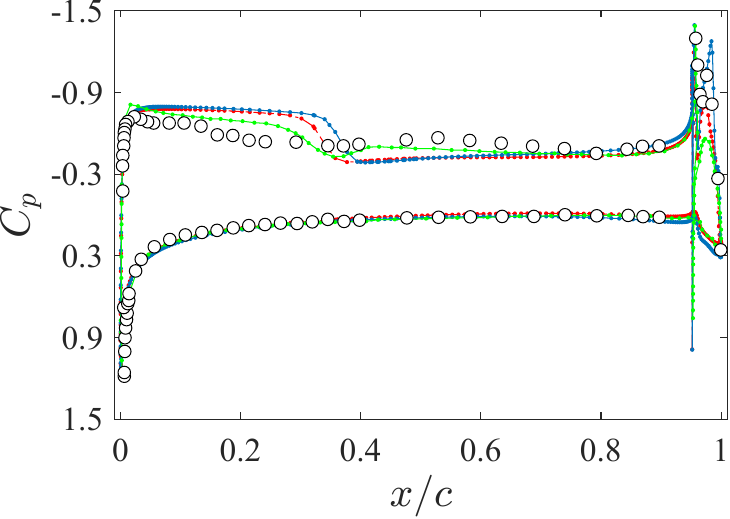}
         \end{subfigure}
         \hfill
         \raisebox{0.82cm}{
         \begin{subfigure}[b]{0.49\textwidth}
             \centering
             \includegraphics[scale=0.975]{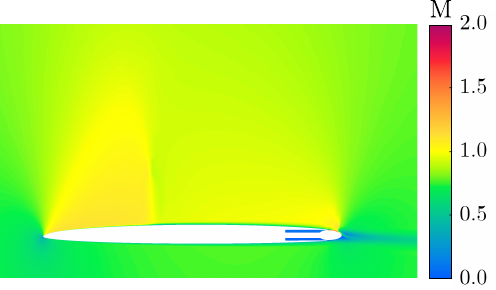}
         \end{subfigure}}
         \caption{jet blowing with $C_\mu = 0.008$}
         \label{fig:TurbValidationAOA3CMU0p008}
     \end{subfigure}
        \caption{Pressure coefficient distribution at mid-span, validated against the TDT experimental data at a free-stream Mach number of 0.8 and angle of attack of $3^{\circ}$, as well as the respective Mach fields using the SA turbulence model.\\ 
        \fbox{\textcolor{navyblue}{\sampleline{} Spalart-Allmaras}, $\;$ \textcolor{red}{\sampleline{dash pattern=on .7em off .2em on .05em off .2em} $k-\omega$ SST}, $\;$ \textcolor{green}{\sampleline{} Chen et al.~\cite{chenNumericalStudyLift2021}}, $\;$ {\stndcirc} TDT tunnel data}}
        \label{fig:TurbValidation}
\end{figure}

Application of circulation control augments the pressure around the airfoil surface, resulting in a wider pressure profile and a shift in the upper surface shockwave further aft, as shown in Fig.~\ref{fig:TurbValidation} which depicts the pressure coefficient distribution for both no blowing ($C_\mu = 0$) and with jet blowing ($C_\mu = 0.008$) as well as the Mach fields of each via the SA model. The calculations reproduce the test data sufficiently across the airfoil surface and exceptionally at the trailing-edge region, yet a discrepancy was observed between numerical predictions and experimental data regarding the shockwave position for both turbulence models. This misalignment is observed particularly using $C_\mu = 0.008$, and it is noted that the $k -\omega$ SST model showed a slightly smaller deviation. 

The misalignment has been encountered in similar studies~\cite{forsterNumericalSimulationTransonic2015a,chanTransonicDragReduction2017,liAirfoilGustLoad2020a}, and the results of the current simulations and those of Chen et al~\cite{chenNumericalStudyLift2021} are compared to illustrate this inconsistency across different calculations. Forester et al~\cite{forsterNumericalSimulationTransonic2015a} suggested that its presence could be due to limitations in assuming a fully turbulent boundary layer, since the experimental setup produced a turbulent boundary layer via an epoxy based trip strip located at 5\%c from the leading-edge which was not geometrically modeled. This conclusion however, is not consistently reflected across prior work. With no uncertainty analysis present in the original test report, an equivalent case could be made for inaccuracies due to sensor limitations within the sensitive shock bubble. Nevertheless, offering predominantly accurate predictions, the current study is believed to reproduce the pressure field with sufficient fidelity for further analysis.

\subsection{Steady State Validation}
\label{subsec:steadyState}

\begin{figure}[h]
    \centering
    \includegraphics[scale=0.8]{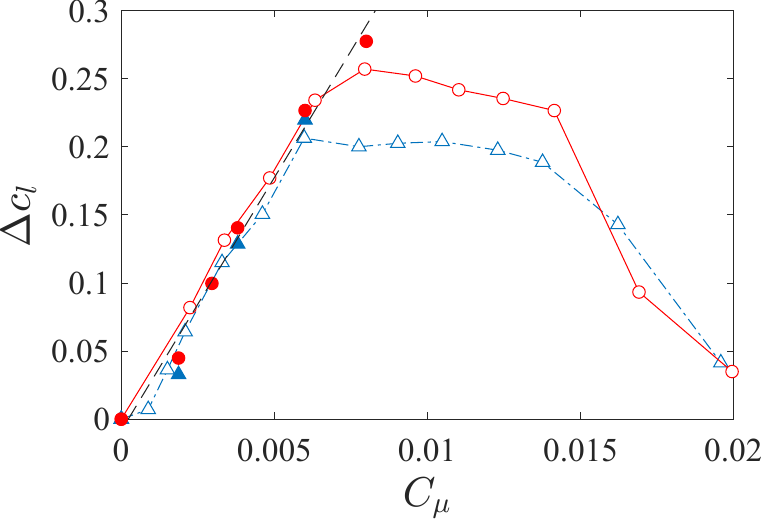}
    \caption{Effect of momentum addition on the mid-span lift enhancement at a free-stream Mach number of 0.8 and an angle of attack of $3^{\circ}$ and $0^{\circ}$.\\
    \fbox{\textcolor{red}{\redcircfilled $\;$ CFD results $\alpha=3^\circ$},  \textcolor{red}{\samplelineShort{}\redcirc\textcolor{red}{\samplelineShort{}} $\;$ TDT tunnel data $\alpha=3^\circ$},
    $\;$ \textcolor{navyblue}{ $\color{navyblue}{\blacktriangle}$ $\;$ CFD results $\alpha=0^\circ$}, \textcolor{navyblue}{\samplelineShort{dashed}$\color{navyblue}{\triangle}$\textcolor{navyblue}{\samplelineShort{dashed}} $\;$ TDT tunnel data  $\alpha=0^\circ$}}}
    \label{fig:CLandAugVSCMU}
\end{figure}

From an integral perspective, the lift enhancement, $\Delta c_l$ is an important parameter, as it isolates the nominal lift of the wing from the lift gained via CC, dictating the efficacy of the of the jet injection on the external aerodynamics. Validation of this feature is presented in Fig.~\ref{fig:CLandAugVSCMU} which illustrates the variation of the lift enhancement as functions of the momentum coefficient at an angles of attack of $ 3^{\circ}$ and $0^\circ$ and a freestream Mach number of $0.8$.

Accurate prediction is achieved compared with the TDT data, especially in the linear region region of the momentum coefficient (\( 0\leq C_\mu \leq 0.006 \)) with a root mean square error (RMSE) of 1.21\% and 2.75\% for $\alpha = 0^\circ,3^\circ$ respectively. The momentum coefficient value of 0.008 was the largest validated in steady-state with a relative error of 7\%, as the augmentation trend trend shifts from linear to a an almost constant behavior with increasing jet strength. 

When the momentum input increases beyond a certain threshold ($C_\mu = 0.0137$), the jet loses its authority, detaches from the surface, and the lift enhancement rapidly declines towards zero with increasing jet strength, preceded by a plateau. In the plateau region (at momentum inputs of \( C_\mu > 0.008 \)), attempts to achieve a steady-state solution revealed unsteady behavior in both the aerodynamic coefficients and the residuals. Similar observations were made in other studies~\cite{forsterNumericalSimulationTransonic2015a, liAirfoilGustLoad2020a}, although no detailed explanation was provided for this phenomenon.

\subsection{Unsteady Validation}

To evaluate the unsteadiness at the plateau, second order Unsteady Reynolds-Averaged Navier-Stokes (URANS) simulations were performed at an angle of attack of $3^\circ$ and a momentum coefficient of 0.01, corresponding to the first indication of unsteady behavior. The results were verified through a time step sensitivity study, which considered three different time increments: $\Delta t = (1\times 10^{-2}, 5\times 10^{-3}, 1\times 10^{-3}) \; \mathrm{(ms)}$, equating to $(128, 256, 1500)$ iterations per cycle, respectively. Convergence was assessed based on a three-order-of-magnitude reduction in residuals per time step, as well as the stabilization of the lift coefficient's amplitude and frequency, determined via a one-sided Fast Fourier Transform (FFT). Among the tested time increments, $\Delta t = 5\times10^{-3} \; \mathrm{(ms)}$ was deemed sufficient for convergence, yielding a lift coefficient amplitude of 0.032 and a reduced frequency $\kappa_\mathrm{jet} = f_{\mathrm{jet}}r_e/\nu_\mathrm{jet}$ of 0.105. Here, $f_{\mathrm{jet}}$ denotes the jet periodic detachment's natural frequency. 

\begin{figure}[h]
     \centering
     \begin{subfigure}[b]{0.49\textwidth}
         \centering
         \includegraphics[scale=0.6]{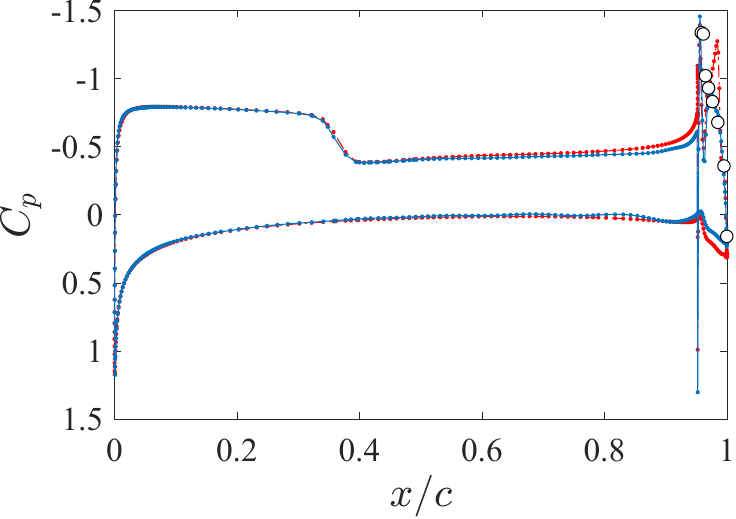}
         \caption{wing}
         \label{}
     \end{subfigure}
     \hfill
     \begin{subfigure}[b]{0.49\textwidth}
         \centering
         \includegraphics[scale=0.6]{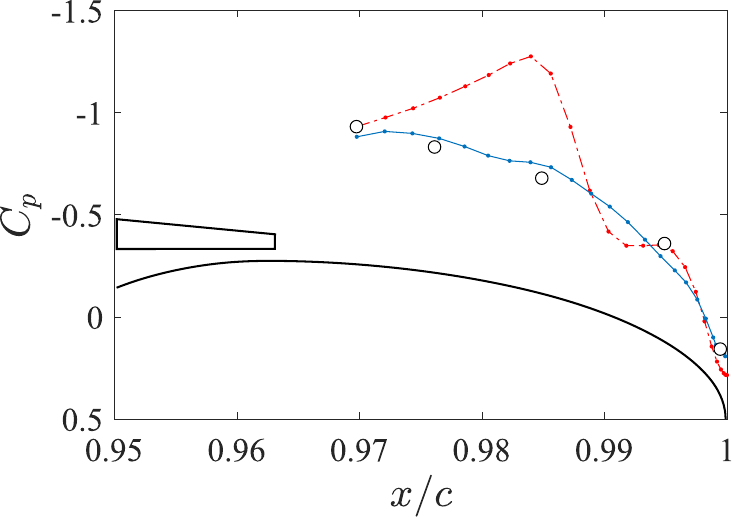}
         \caption{trailing-edge}
         \label{}
     \end{subfigure}
        \caption{Pressure coefficient distributions at mid-span, at a free-stream Mach number of 0.8 and angle of attack of $3^{\circ}$ of the time ensemble CFD data and the steady state data as compared with the TDT experimental result.\\
        \fbox{\textcolor{navyblue}{\sampleline{} time ensemble at $C_\mu = 0.01$}, $\;$ \textcolor{red}{\sampleline{dash pattern=on .7em off .2em on .05em off .2em} $C_\mu = 0.008$}, $\;$ {\stndcirc} TDT tunnel data at $C_\mu = 0.011$}}
        \label{fig:URANSEnsamblePressureDistribution}
\end{figure}

Although no transient data from the TDT experiment was available, a quantitative comparison can nonetheless be performed. This is achieved by computing an ensemble average of the CFD solution over one full cycle and comparing it to the frequency-averaged data provided in the report. Fig.~\ref{fig:URANSEnsamblePressureDistribution} presents the time-averaged CFD pressure distribution at $C_\mu = 0.01$ and the steady value of $C_\mu = 0.008$ alongside the experimental measurements taken near the trailing-edge.

A clear similarity is observed up to the trailing-edge ($0 \leq x/c \leq 0.95$) depicted in Fig.~\ref{fig:URANSEnsamblePressureDistribution}(a) between the steady-state solution at $C_\mu = 0.008$ and the time ensemble at $C_\mu = 0.01$, where the pressure distributions align closely. Notably, in contrast to the behavior at lower jet momentum values ($C_\mu \leq 0.008$), increasing the momentum to the onset of aerodynamic instability does not alter the position of the upper surface shock wave ($0 \leq x/c \leq 0.4$). This suggests that when ejecting a jet of this value, circulation control can no longer augment the upper surface pressure distribution upstream of the trailing-edge. At the trailing-edge region ($0.95 \leq x/c \leq 1$) depicted in Fig.~\ref{fig:URANSEnsamblePressureDistribution}(b) , the pressure peak present at ($0.97 \leq x/c \leq 0.99$) using $C_\mu = 0.008$ is absent at $C_\mu = 0.01$. As the primary feature distinguishing the distributions, its disappearance must pertain to the onset of jet detachment.

\begin{figure}[h]
    \centering
        \begin{subfigure}[b]{0.99\textwidth}
            \centering
            \includegraphics[scale=1]{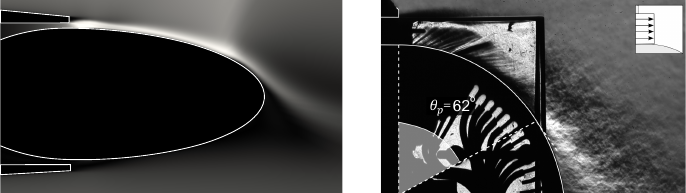}
            \caption{jet attachment}
        \end{subfigure}
    \hfill
    \vspace{1em}
    \hspace*{-0.2cm}
        \begin{subfigure}[b]{0.99\textwidth}
            \centering
            \includegraphics[scale=1]{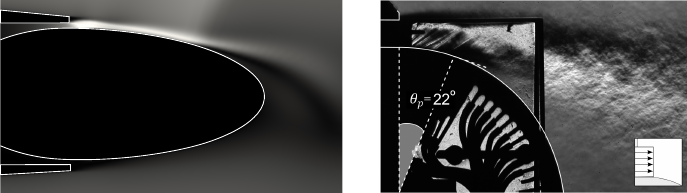}
            \caption{jet detachment}
        \end{subfigure}
    \caption{Jet bi-stability obtained from the URANS calculations and compared to the similar case of bi-stability at quiescent air \cite{jegede2016dual}.}
    \label{fig:URANSvQACompare}
\end{figure}

Further analysis of the flow-field behavior near the trailing-edge indicates that the observed unsteadiness results from the periodic detachment of the jet, as shown in Fig.~\ref{fig:URANSvQACompare} for both the detachment and reattachment phases. The resulting bi-stability closely resembles the well-substantiated periodic separation observed in experimental studies of supersonic jet ejection into quiescent air around a circular trailing-edge~\cite{jegede2016dual,robertson2017influence}, also illustrated in Fig.~\ref{fig:URANSvQACompare}. This similarity is further supported by their use of a comparable nozzle pressure ratio of 3.6, relative to 3.96 in the present study. Therefore, it is reasonable to deduce that the effect observed here represents the same form of bi-stability. However, unlike the quiescent air experiments, the current phenomenon is constrained by the presence of the trailing-edge shockwave due to the elliptical geometry and transonic freestream, which terminates the detachment rather than allowing it to persist until natural equilibrium is reached.

%% file: sections/results.tex
\subsection{Unsteady Flow-Field Analysis}
\label{subsec:unsteady}
\input{sections/unsteady}

\subsection{Dynamic Mode Decomposition}
\label{subsec:DMD}
\input{sections/dmd}

%% file: sections/unsteady.tex
The previous section served primarily to establish the physical validity of the results and to reduce numerical uncertainties. The accurate prediction of both global aerodynamic coefficients and pressure distributions at key locations, together with the clear correspondence of the observed phenomenon to experimentally established jet bi-stability, provides strong confidence that the subsequent exploration of the effect rests on firm physical grounds. 

\begin{figure}[h]
     \centering
        \includegraphics[width=0.98\linewidth]{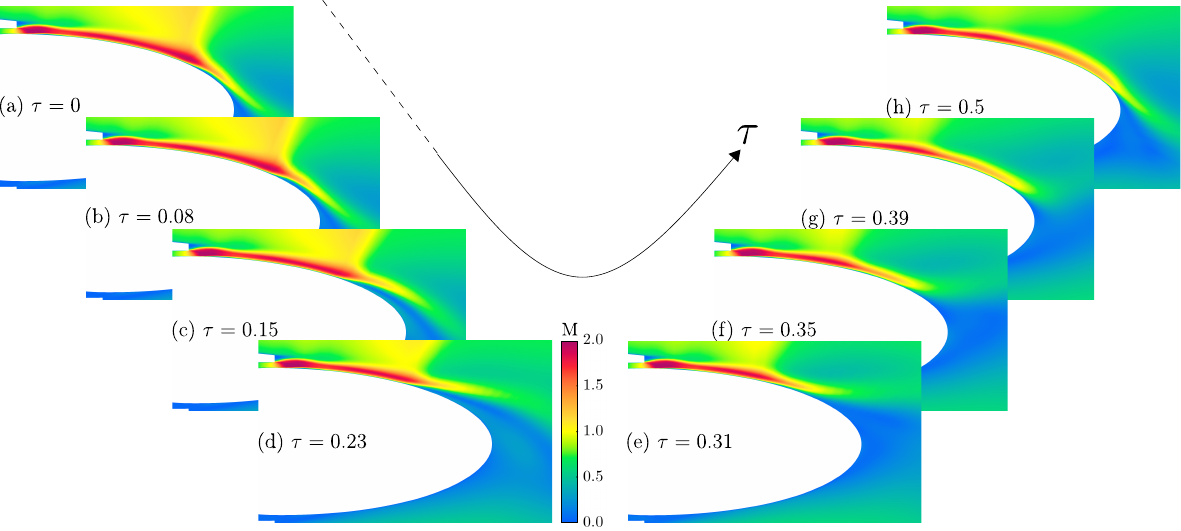}
        \caption{Mach fields of the mid-span depicting the jet bi-stability. With a momentum coefficient of 0.01 at angle of attack of $3^{\circ}$, and free-stream Mach of 0.8. }
        \label{fig:JBContour}
\end{figure}

Analysis of the bi-stability begins with the investigation of its complete cycle in Fig.~\ref{fig:JBContour}. The figure presents Mach fields over a portion of a single detachment cycle at various normalized time steps, defined as $\tau = (t - t_{c,0} / t_c$. Here, $t$ represents the physical flow time, $t_c$ denotes the time of a single cycle and the initial time of the detachment cycle is $t_{c,0}$. Each image in the figure thus represents the relative temporal position of the Mach fields within the cycle.

Starting from Fig.~\ref{fig:JBContour}a ($\tau = 0$), which corresponds to the maximum lift coefficient ($c_l = 0.616$), the jet gradually begins to detach from the surface. This progression is observed in Figs.~\ref{fig:JBContour}(b-d), culminating in full detachment at the minimum lift coefficient ($c_l = 0.581$ at $\tau = 0.23$). Reattachment initiates at $\tau = 0.31$, during which the lift coefficient gradually increases (Figs.~\ref{fig:JBContour}(e-h)), ultimately returning to its peak value, marking the completion of the cycle ($\tau \geq 0.5$).

A key feature of this phenomenon is the interaction between the jet and the trailing-edge shockwave, which weakens as the cycle progresses, as well as the interaction between the jet and the underlying boundary-layer. At $\tau = 0$, and as previously observed in the ensemble data, the unsteady field lacks the pressure peak or separation bubble previously discussed (see Sec.~\ref{sec:numericsNgrid}) for momentum coefficients at the onset of periodic separation ($C_\mu \geq 0.008$). Typically, the separation bubble facilitates mixing between the free-stream momentum and the boundary-layer, thereby delaying detachment~\cite{greenblatt2000control}. Consequently, its absence can lead to premature separation. The high-velocity pushes the separation bubble closer to the tip of trailing-edge, preventing its stable formation.

\begin{figure}[h]
     \centering
     \begin{subfigure}[b]{0.49\textwidth}
         \centering
         \includegraphics[scale=0.6]{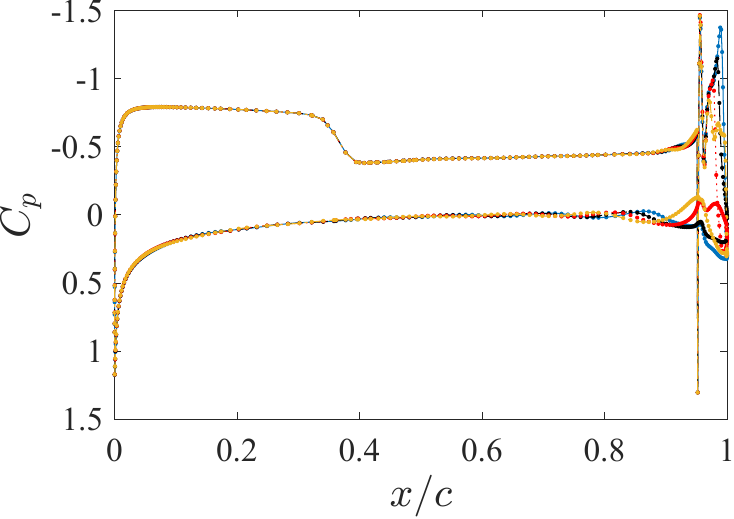}
         \caption{wing}
         \label{}
     \end{subfigure}
     \hfill
     \begin{subfigure}[b]{0.49\textwidth}
         \centering
         \includegraphics[scale=0.6]{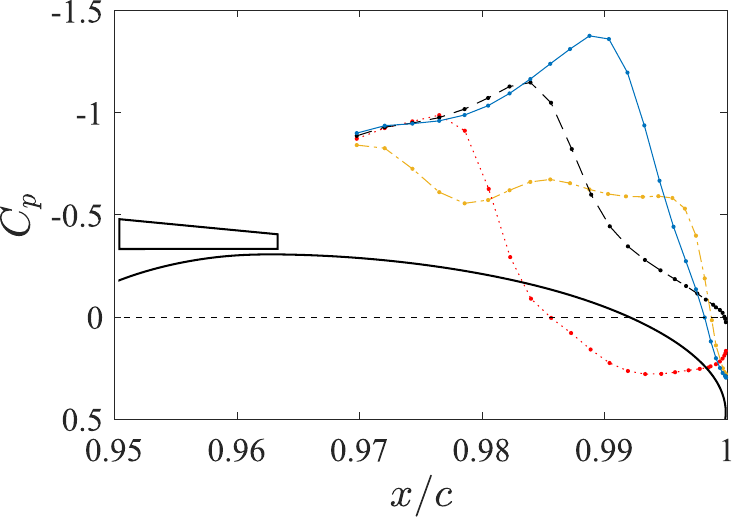}
         \caption{trailing-edge}
         \label{}
     \end{subfigure}
        \caption{Pressure coefficient distribution at mid-span for different parts of the cycle, at a free-stream Mach number of 0.8, angle of attack of $3^{\circ}$ and a momentum coefficient of 0.01.\\ 
        \fbox{\textcolor{navyblue}{\sampleline{} $\tau = 0$}, $\;$ \textcolor{black}{\sampleline{dashed} $\tau = 0.15$}
        , $\;$ \textcolor{red}{\sampleline{dotted} $\tau = 0.31$}, $\;$ \textcolor{dandelion}{\sampleline{dash pattern=on .7em off .2em on .05em off .2em} $\tau = 0.5$}}}
        \label{fig:URANSPressureDistribution}
\end{figure}

To further delineate on the observation, the pressure data explicitly analyzed in Fig.~\ref{fig:URANSPressureDistribution} which depicts the pressure distribution at various stages of the cycle. In Fig.~\ref{fig:URANSPressureDistribution}(b), the absence of the bubble at $\tau = 0$ manifests as a sharp, linear drop in the pressure coefficient within the range $0.99 \leq x/c \leq 1$. As the cycle progresses, this drop becomes more gradual, as seen at $\tau = 0.15$. This change in slope arises from the jet’s inability to sustain a stable separation bubble given the current trailing-edge length, thus supporting the initial detachment caused by the downstream displacement of the bubble. A potential approach to mitigate this would be to extend the trailing-edge length, thereby promoting the formation of a stable bubble. However, this comes at the expense of a flatter pressure distribution curve, which reduces lift enhancement. At $\tau = 0.31$, when the reattachment phase begins, the pressure distribution near $x/c = 0.985$ gradually becomes positive, indicating an adverse pressure gradient acting on the flow, a characteristic signature of flow separation. This unfavorable gradient diminishes as the cycle reaches $\tau = 0.5$, where suction is re-established and a relatively flat pressure distribution occurs, corresponding to reduced lift and the attaining of bi-stability. Despite clear modifications to the pressure field near the trailing-edge caused by the periodic detachment, the upstream surface pressure distribution, depicted in Fig.~\ref{fig:URANSPressureDistribution}(a) remains relatively constant throughout the cycle, reinforcing the decoupling of the bi-stability from the upper-surface shockwave.

The unsteady solution, although underlines the initial detachment as the incapability of the flow to maintain a stable bubble, does not clarify why periodicity is maintained rather than leading to complete jet detachment from the trailing-edge. Furthermore RANS computations do not directly resolve the turbulent eddies responsible for mixing, making it difficult to study the bi-stability phenomenon from the perspective of the separation bubble itself. However, previous studies on periodic mechanisms at transonic speeds~\cite{santana2024analysis, poplingher2019modal, liu2018analysis} (see Sec.~\ref{sec:introduction}) indicate that the harmonic motion is primarily governed by pressure propagation. Moreover, since circulation control (CC) is predominantly a pressure-driven phenomenon~\cite{greenblattFlowControlUnmanned2022}, further insights into the bi-stability can be obtained by analyzing the problem from a pressure-based perspective. As a result of the pressure dominance, the coupling between a primarily pressure-driven effect and periodic motion suggests that this phenomenon can be effectively analyzed using a dynamics-based approach, such as data-driven methods like Dynamic Mode Decomposition (DMD). DMD extracts the spatial-temporal modes of the flow field while relating them to the system’s dynamic quantities. In this context DMD can provide deeper insights into the phenomenon.

%% file: sections/dmd.tex
\input{sections/dmd_introduction}

\subsubsection{Modal Spatial Distribution}
\label{subsubsec:dmdSpatial}
\input{sections/dmd_distribution}

\subsubsection{Mode Time Development}
\label{subsubsec:dmdHist}
\input{sections/dmd_history}

\subsubsection{Pressure Field Reconstruction}
\label{subsubsec:dmdRecon}
\input{sections/dmd_reconstruction}

%% file: sections/dmd_introduction.tex
The modal analysis of the bi-stability is performed using the Dynamic Mode Decomposition (DMD) algorithm. DMD is a data-driven approach for analyzing flow-fields, combining elements of Proper Orthogonal Decomposition (POD) and the Fourier Transform~\cite{kutz2016dynamic}. Like similar techniques, it utilizes Singular Value Decomposition (SVD), a general dimensionality reduction method designed to identify low-dimensional patterns within data~\cite{brunton2022data}. In the context of the observed unsteadiness, DMD serves as a dimensionality reduction tool to capture and investigate the dominant coherent structures that appear to form the bi-stability as well as whether it can be represented in a reduced form. To achieve this, the mean-reduced pressure field is structured into the matrix $X \in \mathbb{R}^{n \times m}$, where the number of rows, $n$, represents the spatial locations of pressure values, and the number of columns, $m$, corresponds to the discrete time steps.

The formal mathematical derivation~\cite{brunton2022data} states that if \( X \in \mathbb{C}^{n \times m} \) is a matrix where \( n \gg m \), then using Singular Value Decomposition (SVD), \( X \) can be decomposed into the left singular matrix, also known as the POD modes, \( \Psi \in \mathbb{C}^{n\times n} \), which optimally describes \( X \) in a least-squares sense~\cite{rowley2017model}, and the right singular matrix \( V \in \mathbb{C}^{m\times m} \), both of which are unitary, satisfying \( \Psi \Psi^* = \Psi^*\Psi = I_\mathrm{id} \) and \( VV^* = V^*V = I_\mathrm{id} \), where the superscript \( ()^* \) denotes the conjugate transpose and $I_\mathrm{id}$ is the identity matrix. These matrices satisfy the eigenvalue equation \( XV = \Psi\Sigma \), where \( \Sigma \in \mathbb{C}^{n\times m} \) is a diagonal matrix containing the data set's singular values arranged in descending order. To obtain a low-rank approximation, a practical approach involves using the \textit{"economy"} SVD, which removes the zero-valued elements of \( \Sigma \). This leads to the reduced form \( X=\hat{\Psi}\hat{\Sigma}\hat{V}^* \), where \( \hat{\Psi} \in \mathbb{R}^{n \times r} \), \( \hat{\Sigma} \in \mathbb{R}^{r \times r} \), \( \hat{V} \in \mathbb{R}^{r \times m} \), and \( r \leq \min(m,n) \).

The DMD algorithm seeks to determine the matrix \( A \) that propagates the snapshots of \( X \) forward in time, such that \( X' \approx AX \), where \( X' \) represents \( X \) advanced by one time step. In a least-squares sense, \( A \) is obtained by solving the minimization problem \( ||X' - AX||_F \), where \( ||\cdot||_F \) denotes the Frobenius norm. Consequently, the non-linear dynamics of the system are approximated linearly through the matrix \( A \). The spatial distributions \( \Phi \), are the eigenvectors of \( A \), associated with the eigenvalues \( \Lambda \in \mathbb{C} \), where the diagonal elements \( \Lambda_{jj} = \lambda_j \) represent the continuous-time eigenvalues. These eigenvalues are mapped as \( \lambda = \ln(\mu)/\Delta t \), where \( \mu \) denotes the discrete-time eigenvalues of \( A \). Using this relation, the system's linearized modal frequencies and decay rates can be determined by

\begin{equation}
    f = \mathrm{img}(\mathrm{ln}(\mu))/2 \pi \Delta t
\end{equation} 

\begin{equation}
    \xi = \mathrm{real}(\mathrm{ln}(\mu))/\Delta t
\end{equation} 

respectively. For a second-order linear system, the natural frequency \( f_n \) and the damping ratio \( \zeta \) are related to \( \lambda \) through the equation: 

\begin{equation}
\label{eq:DynamicLam}
    \lambda = -2 \pi f_n \zeta + i 2 \pi f_n \sqrt{1- \zeta^2}   
\end{equation}

From this, the natural frequency and damping ratio can be computed as $f_n = |\lambda|/2\pi$ and $\zeta = -\xi/f_n$.
In practice, the modes and frequencies are determined by projecting the matrix \( A \) onto the Proper Orthogonal Decomposition (POD) modes~\cite{kutz2016dynamic}. This is expressed as \( \tilde{A} = \Psi^* A \Psi \), which utilizes the Singular Value Decomposition (SVD), leading to \( \tilde{A} = \Psi^* X' V \Sigma^{-1} \). The matrix \( \tilde{A} \) shares the same eigenvalues as \( A \) and provides a partial approximation of its eigenvectors. Using the eigendecomposition of \( \tilde{A} \), the eigenvectors \( W \) are employed to reconstruct the dynamic modes as $\Phi = X' V \Sigma^{-1} W$. As a result, the continuous-time solution for the pressure field is expressed as:  

\begin{equation}
\label{eq:DMDEq}
    p(t) \approx \sum_{j=1}^r \phi_j e^{\lambda_j t} b_j
\end{equation}  
where \( b = \Phi^{-1} p(t(0)) \) represents the modal amplitude.  

\subsubsection{Preliminary Modal Analysis}
\label{subsubsec:prelim}

\begin{figure}[h]
    \centering
    \includegraphics[scale=0.8]{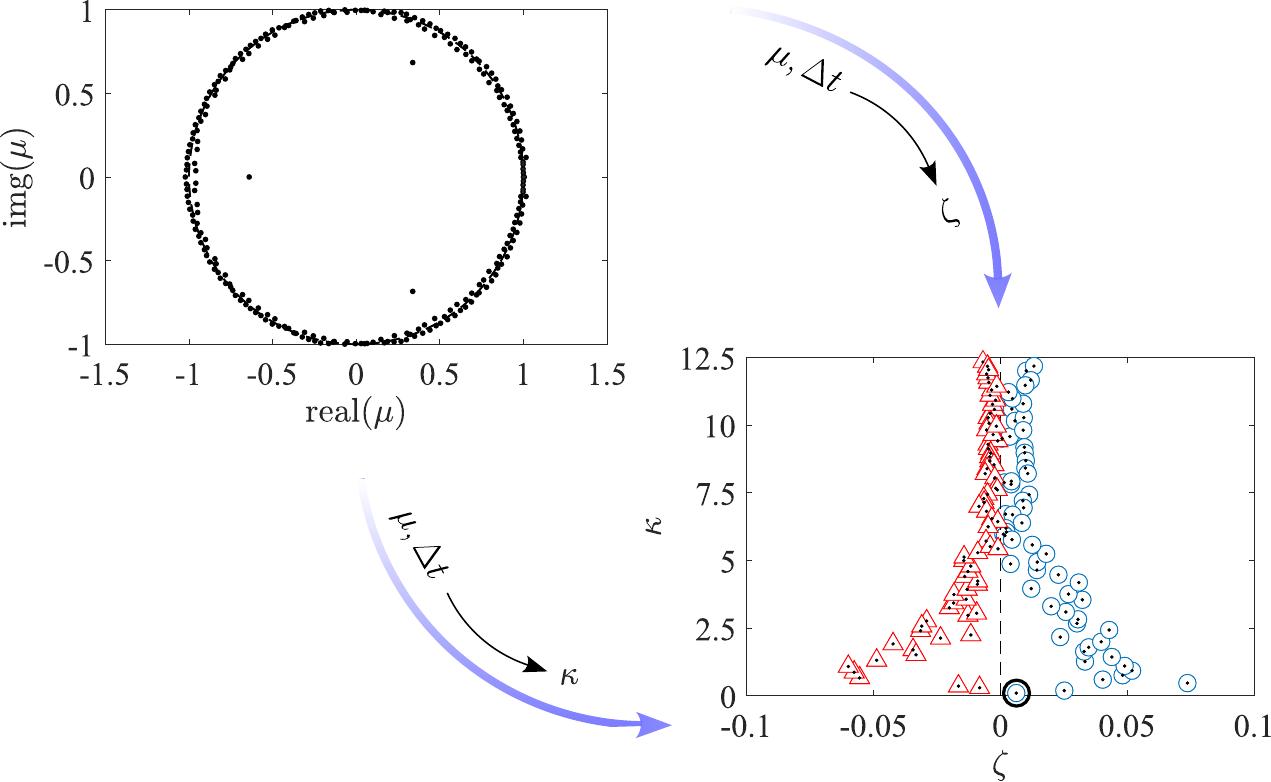}
    \caption{Analysis of the pressure field mode scatter behavior for the modal mapping to the complex plane and their transformation to dynamic properties.\\
     \fbox{\bluecirc \textcolor{navyblue}{$\;$underdamped modes$ \; 0<\zeta < 1$}, \textcolor{red}{ $\normalsize{\triangle} \;$driving modes$ \; \zeta < 0$}}}
    \label{fig:DMDPerlimMapping}
\end{figure}

The DMD process begins by identifying the dominant modes that capture the bi-stability phenomenon. Fig.~\ref{fig:DMDPerlimMapping} shows the eigenvalues of matrix \( A \) plotted on the complex plane for the total 267 snapshots used,  provides insight into the dynamic stability of each mode. The vicinity of the eigenvalues to the unit circle is an indicator of the mode's dynamic stability. Modes that are naturally stable tend closer to the unit circle~\cite{kutz2016dynamic}, while those further from the circle are more unstable and contribute less to the sustained dynamics. To further interpret these modes, they are transformed in terms of their damping ratio \( \zeta \) and reduced frequency \( \kappa \) (see Sec.~\ref{subsec:unsteady}). Red marked points, where \( \zeta < 0 \), represent modes that amplify over time, destabilizing the system due to amplification of the exponential term in~Eq.\ref{eq:DMDEq} and are referred throughout as the "driving modes", while blue marked points with \( 0 \leq \zeta \leq 1 \) indicate "underdamped modes", that decay or oscillate stably. Because the bi-stability is characterized by a low-frequency periodic motion, the relevant mode is expected to lie at the minimum reduced frequency, i.e., \( \kappa_{\mathrm{jet}} = \kappa_{\mathrm{min}} \), with a damping ratio tending toward zero. In Fig.~\ref{fig:DMDPerlimMapping}(a), the eigenvalue \( \mu_{\mathrm{jet}} = 0.9995 + 0.0244i \) corresponds to a frequency of \( f_{\mathrm{jet}} = 777.86 \, \mathrm{Hz} \) or \( \kappa = 0.096 \), meeting these criteria. This mode is marked with a black circle and aligns well with the dominant frequency identified in the prior FFT analysis.

To further assess their relevance, the contribution of these modes to the bi-stability, is examined based on their influence on the flow-field. Although a variety of parameters which analyze accumulating contributions are present in the literature, this study utilizes the modal influence~\cite{kou2017improved}. It accounts for both the modal amplitude \( b \) and the decay rate \( \xi \) adding weight to weakly energetic effects through their contribution to the exponential term. The modal influence, $I$ is defined as

\begin{equation}
\label{eq:modalInfluence}
    I_r = \sum_{j=1}^r b_j ||\lambda_j||   
\end{equation}

\begin{figure}[h]
     \centering
     {\includegraphics[scale=0.8]{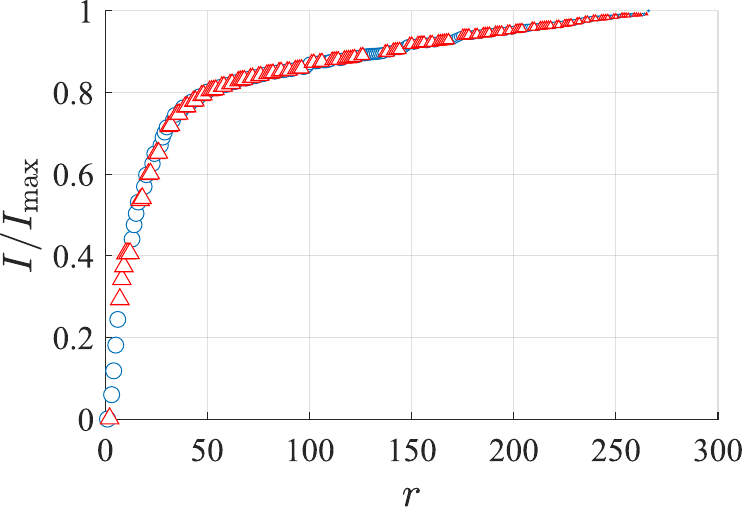}}
     \caption{Contribution of the accumulating modal influence relative to the total influence with increasing mode number.\\
     \fbox{\bluecirc \textcolor{navyblue}{$\;$underdamped modes$ \; 0<\zeta < 1$}, \textcolor{red}{ $\normalsize{\triangle} \;$driving modes$ \; \zeta < 0$}}}
     \label{fig:DMDPerlimInf}
\end{figure}

\begin{figure}[h]
     \centering
     {\includegraphics[scale=0.8]{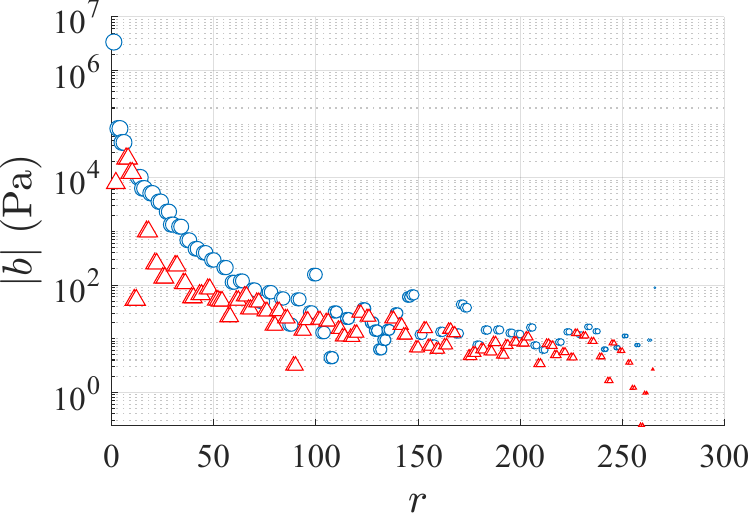}}
     \caption{Contribution of the modal amplitude with increasing mode number.\\
     \fbox{\bluecirc \textcolor{navyblue}{$\;$underdamped modes$ \; 0<\zeta < 1$}, \textcolor{red}{ $\normalsize{\triangle} \;$driving modes$ \; \zeta < 0$}}}
     \label{fig:DMDPerlimAmp}
\end{figure}

Fig.~\ref{fig:DMDPerlimInf} illustrates the modal normalized influence, $I/I_\mathrm{max}$ as a function of mode number, while Fig.~\ref{fig:DMDPerlimAmp} shows the amplitude of both. As observed in Fig.~\ref{fig:DMDPerlimInf}, the first mode (\( r = 1 \)), corresponding to the time-averaged pressure field as well as the second mode ($r = 2$), exhibit almost no influence on the bi-stability, unlike subsequent modes. Notably, 24.5\% of the total modal influence is concentrated in less than 2\% of the underdamped modes, namely the two modal pairs at \( r = 3,4 \) and \( r = 5,6 \). Although a full reconstruction of the flow field would require more modes, these pairs contribution is significant to the dynamics.

In Fig.~\ref{fig:DMDPerlimAmp}, the amplitude of each mode is shown to decay with increasing mode number. The dominant underdamped modes up to the second modal pair (\( r = 4,5 \)) display an order-of-magnitude difference in amplitude compared with later modes. The driving modes, while present, are of lower magnitude and exhibit more erratic behavior. A clear contrast in trends is seen between the underdamped and driving modes, with the former following a more consistent decay. It is thus observed that the impact of the first modal pairs is substantial. Consequently, constructing a reduced model of the bi-stability requires examining these modes directly in relation to the rest and ascertaining their influence on the pressure field.

%% file: sections/dmd_distribution.tex
\begin{figure}[h]
     \centering
     \begin{subfigure}[b]{0.49\textwidth}
         \centering
         \includegraphics[scale=0.54]{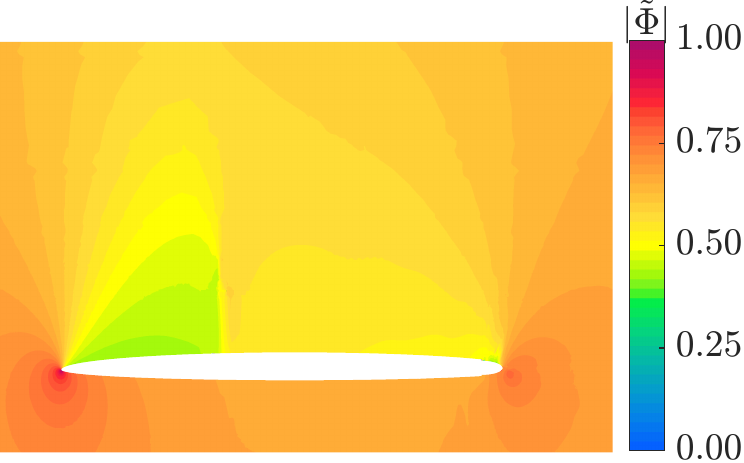}
         \caption{$r=1$}
         
     \end{subfigure}
     \hfill
     \begin{subfigure}[b]{0.49\textwidth}
         \centering
         \includegraphics[scale=0.54]{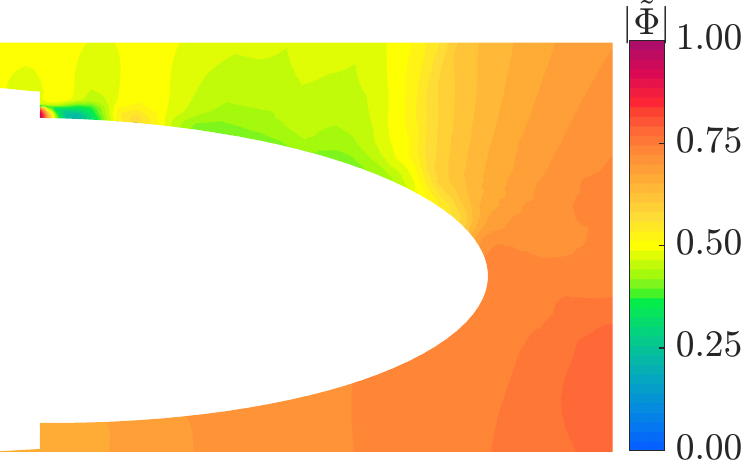}
         \caption{$r=1$}
         
     \end{subfigure}
     \hfill
     \begin{subfigure}[b]{0.49\textwidth}
         \centering
         \includegraphics[scale=0.54]{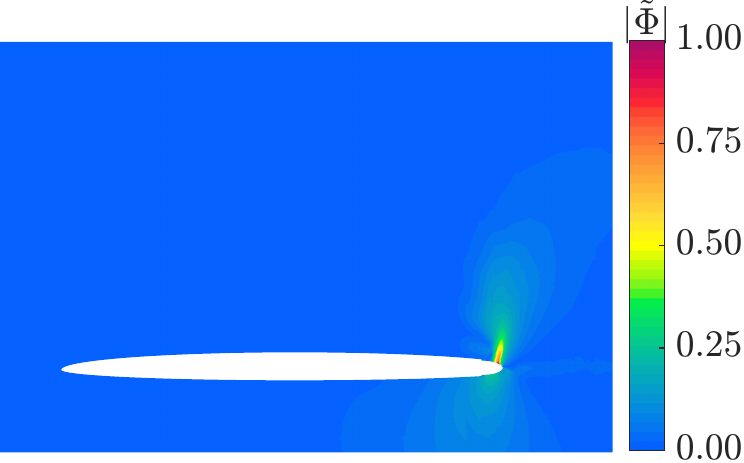}
         \caption{$r=3,4$}
         
     \end{subfigure}
     \hfill
     \begin{subfigure}[b]{0.49\textwidth}
         \centering
         \includegraphics[scale=0.54]{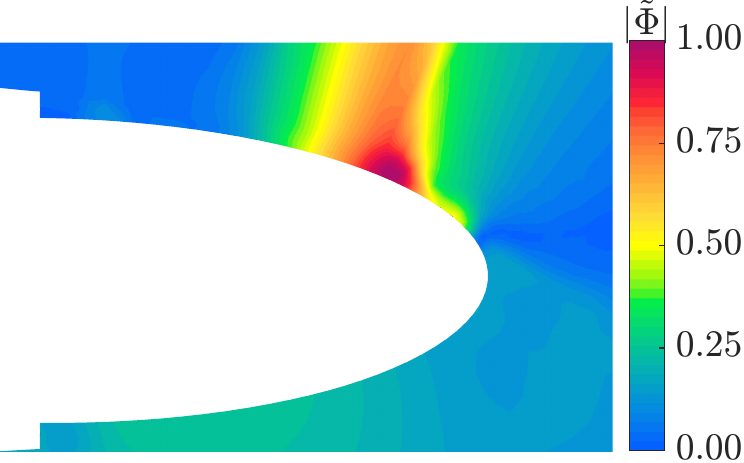}
         \caption{$r=3,4$}
         
     \end{subfigure}
        \caption{Spatial distribution field of the first steady and unsteady modes.}
        \label{fig:DMDModes134Compare}
\end{figure}

Figure ~\ref{fig:DMDModes134Compare} shows the normalized spatial amplitude \( |\tilde{\Phi}| = |\Phi/\mathrm{max}(\Phi)| \) of both the first mode, which corresponds to the mean pressure field, and the first modal pair (\(r = 3, 4\)), which not only relates to the bi-stability frequency but also has the highest modal amplitude among the buffet modes. The second mode $r=2$ as discussed previously, possessed almost no modal influence, and did not contain additional information from its spatial distribution which was found to be relevant to the bi-stability. As previously noted, increasing the jet momentum while the bi-stability persists, does not change the location of the upper surface shockwave. From Fig.~\ref{fig:DMDModes134Compare}(a,c), it can be seen that while the upper surface shock is present in the first mode, it is absent in the modal pair and in subsequent modes. This absence, which was outlined in the URANS results, is due to the jet's inability to continue augmenting the circulation of the wing and as a result, cannot further shift the shock's position. Therefore, it is understood that bi-stability marks the end of circulation control, even prior to complete jet separation. Also noted is a pressure structure seen on the surface of the trailing-edge at \( r = 3, 4 \) in Fig.~\ref{fig:DMDModes134Compare}d, which extends above the surface and contains a pressure core characterized by large gradients from its center, as well as a smaller pressure bubble further downstream. These structures match in form to the interaction of the trailing-edge shock with the unstable separation bubble.

\begin{figure}[h]
     \centering
     \begin{subfigure}[b]{0.3\textwidth}
         \centering
         \includegraphics[scale=0.43]{imgs/Results/DMD/SpatialDist/SDTEView3Mode3.pdf}
         \caption{$r=3,4$}
         
     \end{subfigure}
     \hfill
     \begin{subfigure}[b]{0.3\textwidth}
         \centering
         \includegraphics[scale=0.43]{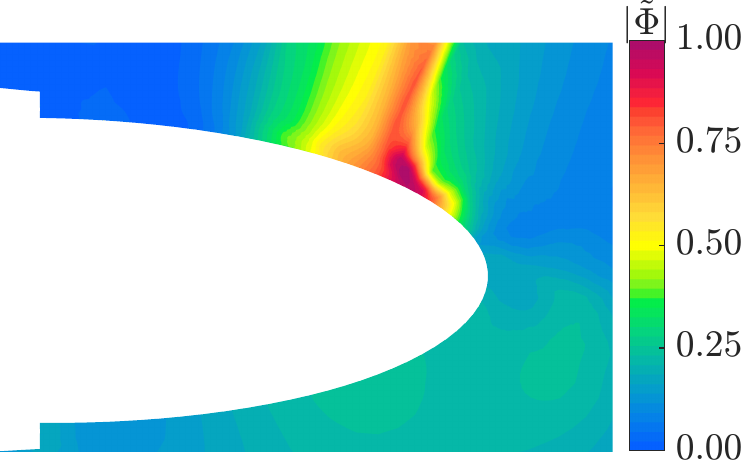}
         \caption{$r=5,6$}
         
     \end{subfigure}
     \hfill
     \begin{subfigure}[b]{0.3\textwidth}
         \centering
         \includegraphics[scale=0.43]{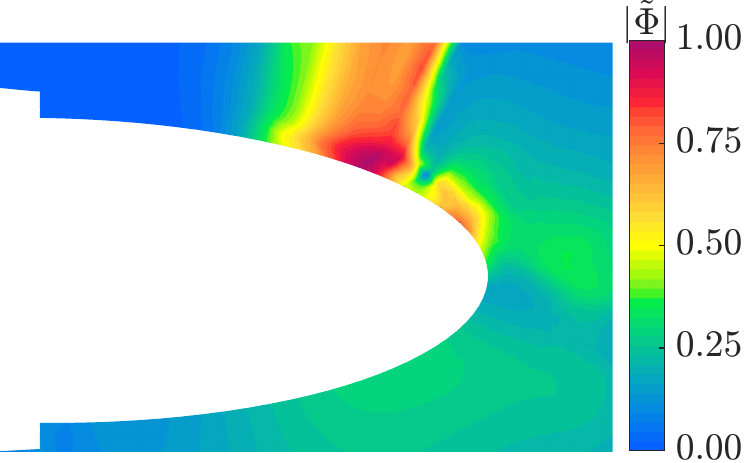}
         \caption{$r=7,8$}
         
     \end{subfigure}
     \hfill
     \begin{subfigure}[b]{0.3\textwidth}
         \centering
         \includegraphics[scale=0.43]{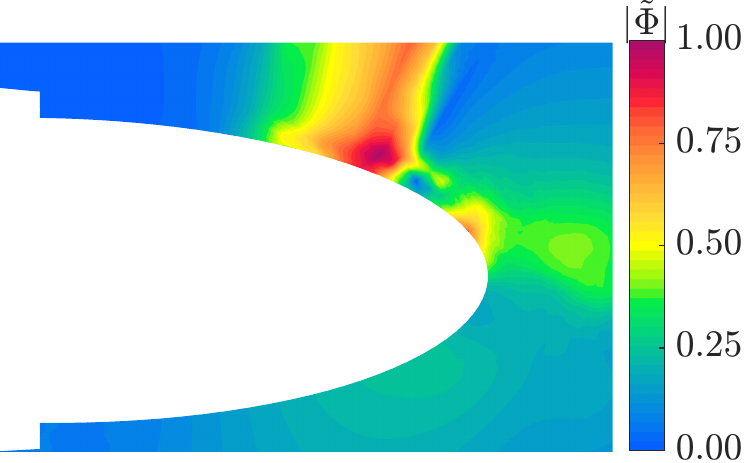}
         \caption{$r=9,10$}
         
     \end{subfigure}
     \hfill
     \begin{subfigure}[b]{0.3\textwidth}
         \centering
         \includegraphics[scale=0.43]{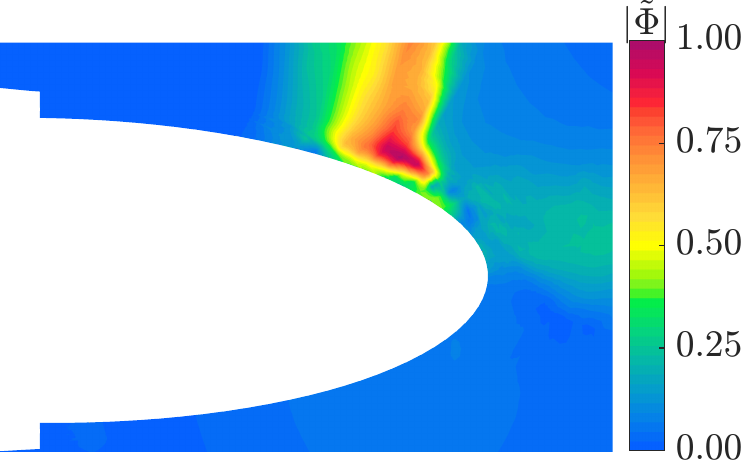}
         \caption{$r=11,12$}
         
     \end{subfigure}
     \hfill
     \begin{subfigure}[b]{0.3\textwidth}
         \centering
         \includegraphics[scale=0.43]{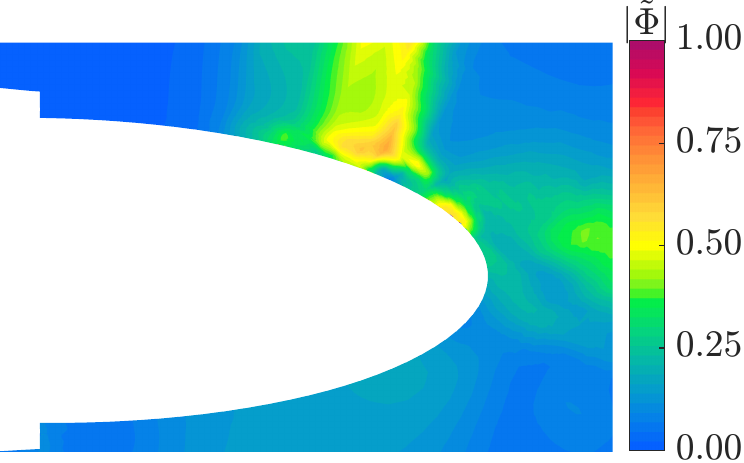}
         \caption{$r=13,14$}
         
     \end{subfigure}
        \caption{Spatial distribution of the first six modal pairs.}
        \label{fig:ModesContour}
\end{figure}

Superseding the latter modes are those shown in Fig.~\ref{fig:ModesContour}. These modal pairs account for 44\% of the modal influence, and as the mode number increases their spatial contribution decreases. The pressure structure first observed at \( r = 3, 4 \) becomes fragmented as the mode number increases, separating into an individual shock structure and a small pressure bubble. Meanwhile, its high-amplitude core steadily rises above the surface of the trailing-edge. The pressure bubble advances along the trailing-edge until it is shed into the wake, while the shock is gradually weakened and detached. When the higher modes are reached \( r = 11, 12 \) in Fig.~\ref{fig:ModesContour}e, the shock structure is completely detached, and the bubble is no longer visible. 

From the perspective of the system's damping characteristics, modes $r = 3 - 6$ as well as $r=13,14$ are underdamped (\( 0 < \zeta < 1 \)) and thus contribute to the stable oscillatory portion of the unsteady flow. In contrast, modes \( r = 7 - 12 \) are classified as driving modes (\( \zeta < 0 \)) and are responsible for destabilizing the system. The difference in spatial structure between these two groups offers insight into this contrast: modes \( r = 3-6 \) preserve the initial pressure configuration, exhibiting only slight displacements of the pressure core as the structure begins to deform. Conversely, modes \( r = 7-12 \) reveal a fragmented pressure field, indicating their role in the breakup process. Consequently, it is this breakup of the pressure structure that introduces destabilizing dynamics into the system.

\begin{figure}[H]
     \centering
     \begin{subfigure}[b]{0.3\textwidth}
         \centering
         \includegraphics[scale=0.43]{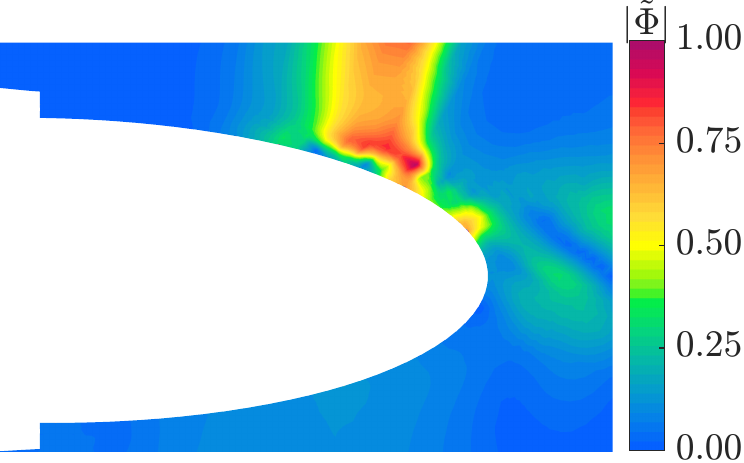}
         \caption{$r=15,16$}
         
     \end{subfigure}
     \hfill
     \begin{subfigure}[b]{0.3\textwidth}
         \centering
         \includegraphics[scale=0.43]{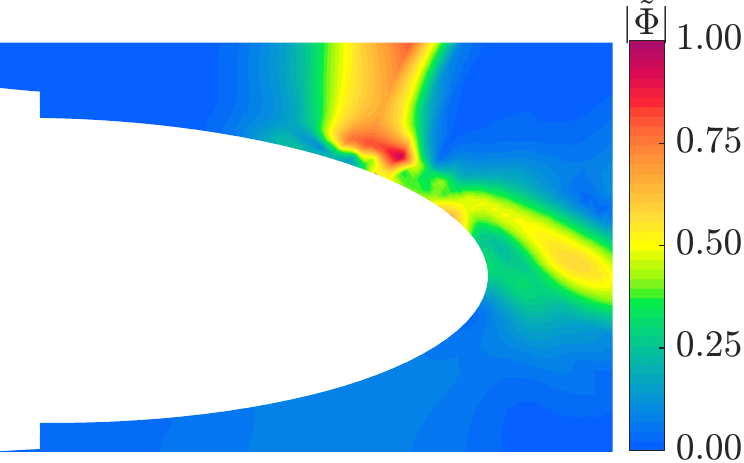}
         \caption{$r=19,20$}
         
     \end{subfigure}
     \hfill
     \begin{subfigure}[b]{0.3\textwidth}
         \centering
         \includegraphics[scale=0.43]{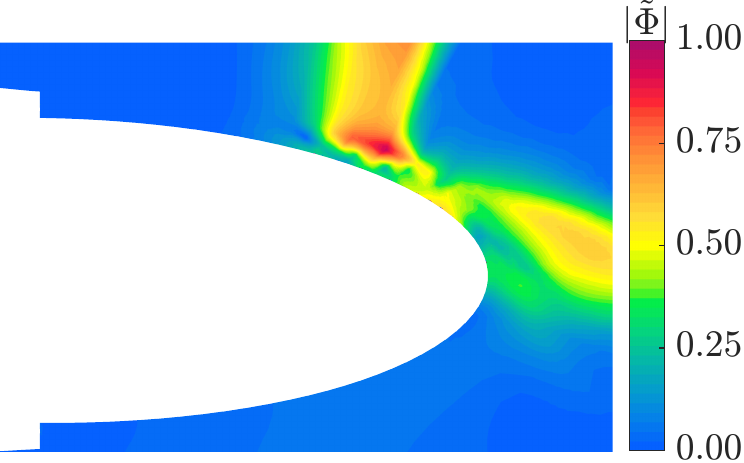}
         \caption{$r=23,24$}
         
     \end{subfigure}
     \hfill
     \begin{subfigure}[b]{0.3\textwidth}
         \centering
         \includegraphics[scale=0.43]{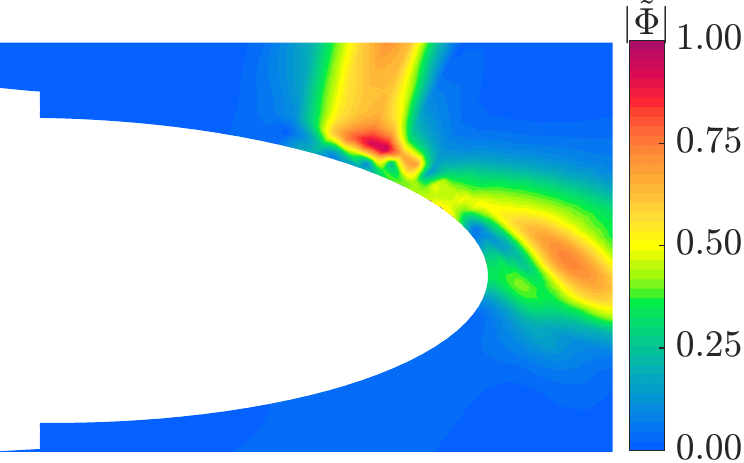}
         \caption{$r=27,28$}
         
     \end{subfigure}
     \hfill
     \begin{subfigure}[b]{0.3\textwidth}
         \centering
         \includegraphics[scale=0.43]{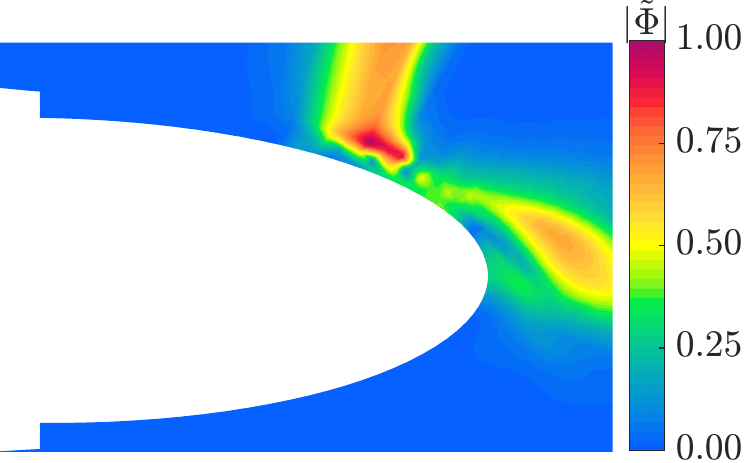}
         \caption{$r=29,30$}
         
     \end{subfigure}
     \hfill
     \begin{subfigure}[b]{0.3\textwidth}
         \centering
         \includegraphics[scale=0.43]{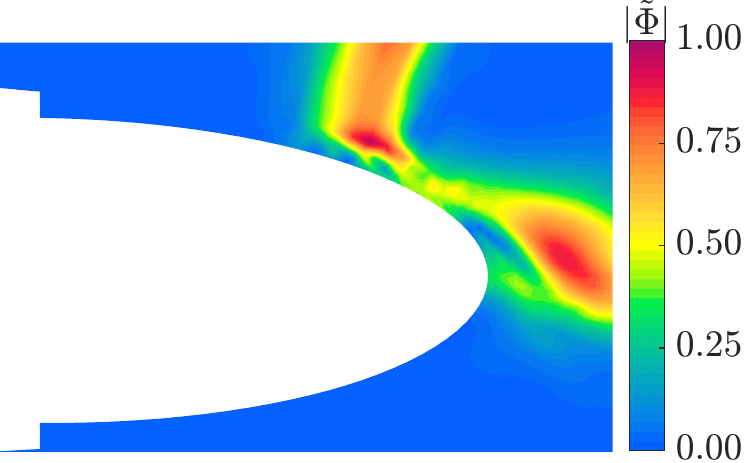}
         \caption{$r=33,34$}
         
     \end{subfigure}
        \caption{Spatial distribution of positive damping modes related to trailing-edge vortex shedding.}
        \label{fig:SheddingModesContour}
\end{figure}

Modes \( r = 13,14 \) are also an underdamped modal pair; however, unlike the earlier underdamped modes, they exhibit substantial dissipation of both the shock and the pressure bubble. This extensive dissipation suggests that beyond a certain mode number, the influence of the bi-stability phenomenon weakens, while other dynamics, such as vortex shedding, begin to dominate. This transition is illustrated in Fig.~\ref{fig:SheddingModesContour}, which presents the spatial structure of the underdamped modes following \( r = 13,14 \). Although these higher-order modes still capture the detached shock seen in \( r = 13,14 \), they differ in the near-wake region, where vortex shedding becomes prominent across all modes. Furthermore, the two-order-of-magnitude difference in amplitude between the lower underdamped modes associated with the initial bi-stability structure (\( r = 3-6 \)) and the higher-order shedding modes highlights a gradual shift from the bi-stability mechanism to vortex shedding. As a result, the latter modes contribute minimally to the bi-stability. The absence of the coherent structures observed in the first and second modal pairs reinforces the conclusion that those early modes are the primary governing mechanism of the bi-stability cycle.

%% file: sections/dmd_history.tex
\begin{figure}[H]
     \centering
        \includegraphics[width=0.98\linewidth]{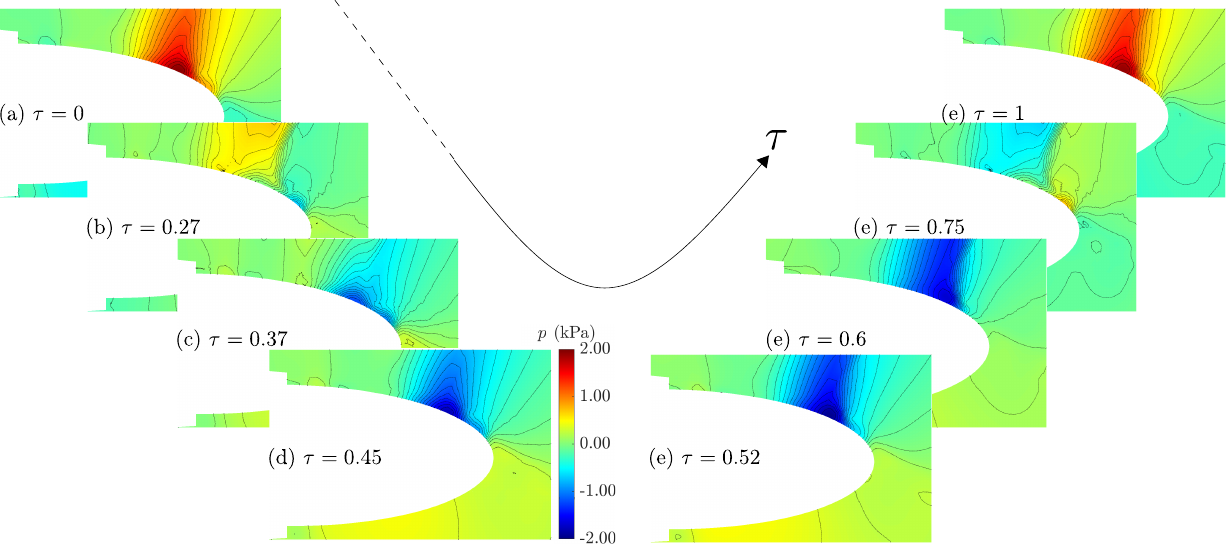}
        \caption{Time history of $p_{3,4}$ (corresponding to $r=3,4$)} depicting the interaction and feedback of the trailing-edge shock wave with the upstream pressure bubble.
        \label{fig:DMDTimeHistory3}
\end{figure}

To delineate on the observation, the time development and contribution to the pressure reconstruction of the underdamped modal pairs, $p_{3,4}$ and $p_{5,6}$ where, \( p_j = b_j \Phi_j e^{\lambda_j t} \), can be examined. Shown in Fig.~\ref{fig:DMDTimeHistory3} is the time development of $p_{3,4}$, for different pressure distributions along the bi-stability cycle. Where \( \tau = 0, 1 \) correspond to the reconstructed maxima of the single mode. During these times, the high amplitude of the shock-bubble structure on the trailing-edge has a positive sign due to the jet attachment. Meanwhile, at \( \tau = 0.45, 0.52 \), the combined structure has a negative sign due to the jet detachment. Additionally, \( \tau = 0.27, 0.75 \) represent the points in the cycle where the trailing-edge pressure structure shifts its phase. At these temporal locations, it is observed that when the phase shifts, the pressure bubble is in anti-phase with the shock formation which subsequently separates from the trailing-edge surface. Thus, the primary mechanism of the bi-stability relies on the feedback from the bubble downstream. As the cycle progresses, pressure waves propagate from the pressure bubble to the shock formation and vice versa, causing an interference that detaches and reattaches the shock to the surface. This finding aligns with the experimental conclusions for the detachment of a supersonic jet at quiescent air \cite{jegede2016dual} where the bubble played a major role in the bi-stable mechanism of attachment and reattachment. Nevertheless, unlike the bi-stability at quiescent air, the absence of a dominant anti-phase with another pressure bubble, solidifies the difference between the the two cases. The presence of the trailing-edge shock plays a major role in the stable periodic motion of the bi-stability.

\begin{figure}[H]
     \centering
        \includegraphics[width=0.98\linewidth]{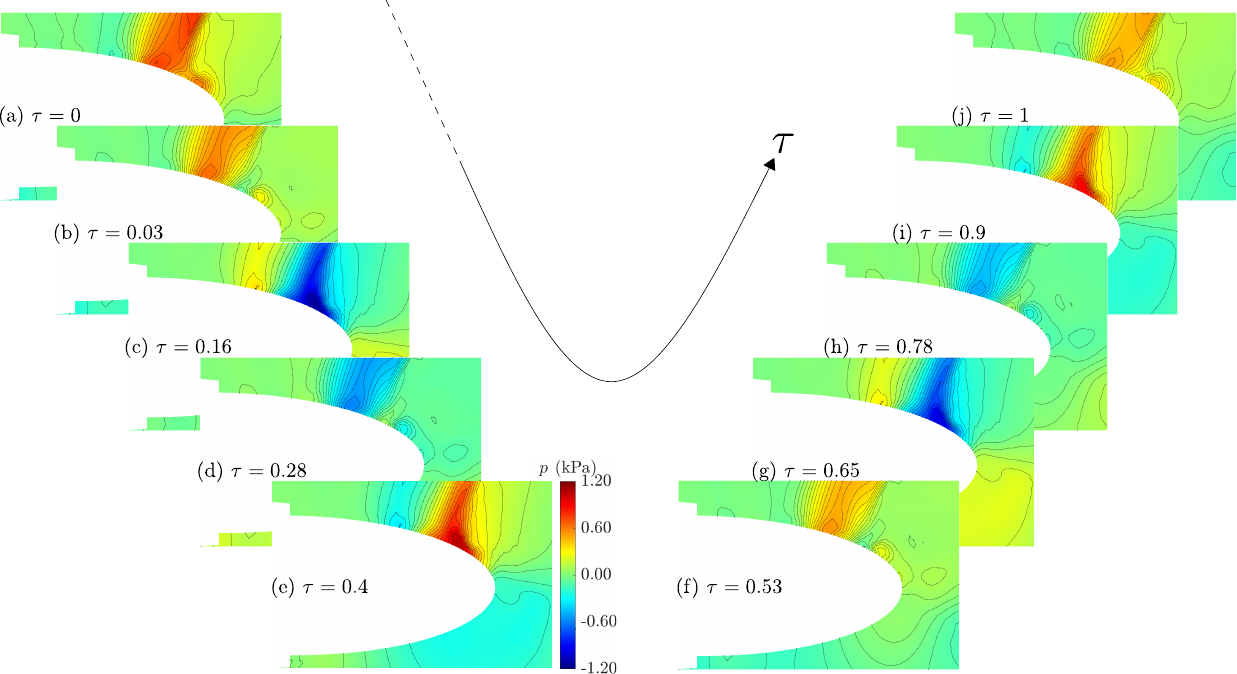}
        \caption{Time history of $p_{5,6}$ (corresponding to $r=5,6$)} depicting the two distinct phase shifts of both the trailing-edge shock and pressure bubble. 
        \label{fig:DMDTimeHistory5}
\end{figure}

Since all modes are linearly correlated, the second modal pair (\( r = 5,6 \)) corresponds to the second harmonic of the primary mode. As illustrated in its pressure reconstruction in Fig.~\ref{fig:DMDTimeHistory5}, two distinct phase shifts can be identified: one at \( \tau = (0.16,\, 0.40,\, 0.65,\, 0.90) \) and another at \( \tau = (0.03,\, 0.28,\, 0.53,\, 0.78) \). The first set reflects the separation of the combined pressure structure observed in \( r = 3,4 \) into two discrete structures, enclosing the pressure bubble and interacting through a feedback mechanism. The second set is more subtle, matching to a similar reconstruction process but focused on the pressure bubble itself, with the shockwave remaining static. Given that the modal pair \( r = 5,6 \)  was previously associated with the reshaping of the initial pressure structure, these phase shifts must correspond to feedback processes responsible for that reshaping. Thus, the second modal pair is not responsible for initiating attachment or detachment, but instead characterizes how the pressure field reorganizes as a result of those processes.

%% file: sections/dmd_reconstruction.tex
The conclusions from prior sections, are used as a foundation for the pressure reconstruction. Since the $r=3,4$ and $r=5,6$ represent the underline detachment, reattachment and pressure redistribution mechanism, if added to the averaged pressure field, should capture the bi-stability mechanism and serve as a reduced order model for the effect. To determine this, the error in prediction is evaluated by the root mean square of the pressure field, $\mathrm{RMSE_p}$ at each cycle time, normalized by the dynamic pressure, to create a representation of the deviation in the pressure coefficient, i.e. $\delta C_p = \mathrm{RMSE_p}/Q_\mathrm{dyn}$. 

\begin{figure}[H]
     \centering
     {\includegraphics[scale=0.8]{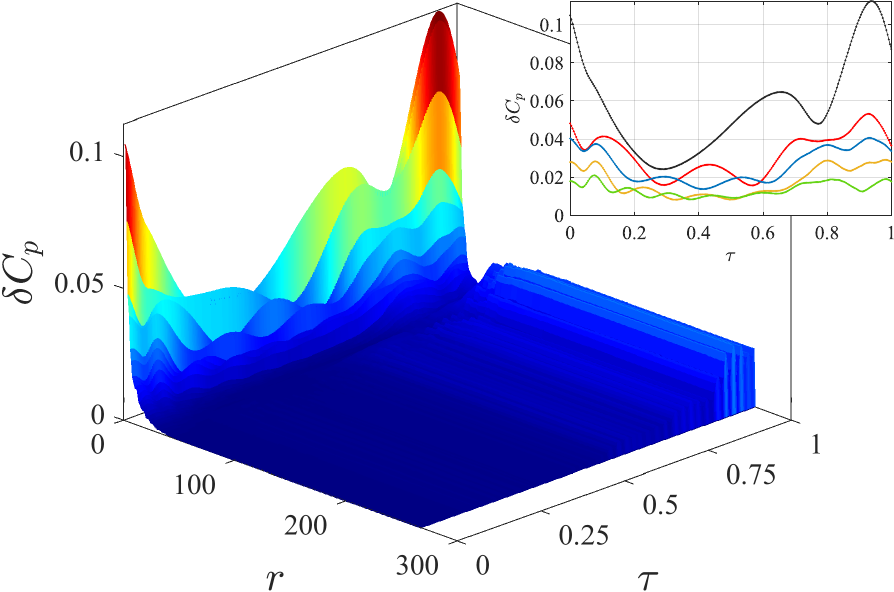}}
     \caption{Accuracy of the pressure field reconstruction for both the total number of modes as well as up to the six most contributing modal pairs.\\
     \fbox{\textcolor{black}{\sampleline{} $r = 1$}, \textcolor{red}{\sampleline{} $r = 1-3$}, \textcolor{navyblue}{\sampleline{} $r = 1-4$}, \textcolor{dandelion}{\sampleline{} $r = 1-5$}, \textcolor{matgreen}{\sampleline{} $r = 1-6$}}}
     \label{fig:DMDReconstruction}
\end{figure}

Figure~\ref{fig:DMDReconstruction} shows the reconstruction error as a function of increasing modal number. The error decreases rapidly as more modes are included. Using only the first mode results in a 10\% error, but the reconstructed signal remains entirely static. In contrast, reconstruction using the first 24 modes reduces the error to 0.5\%, albeit at the cost of including a larger number of terms in the reduced model. This trend aligns with the earlier observation from the modal influence analysis, where most of the flow's dynamic content was carried by a small subset of modes.

Conversely, when additional modes are included (\( r > 24 \)), a persistent error peak emerges near the end of the cycle. The consistent location of this peak, and its appearance only beyond a certain mode count, suggests that a specific dynamic behavior arises at this stage of the cycle that is not adequately captured by the DMD. It may stem from the transition of the flow from separated shear-layer and vortex shedding back to bi-stability. Nevertheless, the maximum 2\% error associated with it remains within typical engineering tolerances.

Figure~\ref{fig:DMDReconstruction} also illustrates this behavior in more detail, showing the temporal evolution of reconstruction error for the lower modes. With only the first mode, the error profile exhibits an harmonic trend, indicating a failure to capture the underlying oscillatory behavior. As more modes are added, the error profile flattens, with only small residual oscillations. When all six significant modes are included, the maximum error drops to approximately 2\%, occurring only near the cycle edges. These results underscore not only the importance of the \( r = 3,4 \) and \( r = 5,6 \) modal pairs, but also the ability of a compact ROM to represent the complex unsteady pressure field and its associated lift dynamics with up to 98\% accuracy.

%% file: sections/conclusions.tex
A numerical validation via RANS was performed for an experimental study investigating the applications of circulation control (CC) on an elliptic airfoil at transonic conditions. The steady-state results aligned well with the experimental study, matching the pressure distribution and integral coefficients.

Upon assessing the contribution of CC at higher momentum coefficients (\( C_\mu = 0.01 \)), unsteadiness was observed, as seen in previous studies, prompting the use of a URANS simulation to investigate the unsteady flow-field. Both a quantitative comparison with the TDT time-averaged data, as well as a visual comparison with the similar effects, revealed an interaction between the high-momentum jet, the trailing-edge shock, and the separation bubble. 

This process caused the bubble to be pushed beyond the trailing-edge tip and thus destabilize the jet. This interaction induced bi-stable oscillations in the lift coefficient. Further investigation revealed that the bi-stability dynamics were largely decoupled from the upper surface shock. This decoupling resulted from the inability of the jet to further augment the wing’s circulation, preventing the wing from conforming to super-circulation.

To both understand the feedback mechanism necessary to create persistent oscillations, as well as deduce if the bi-stability can be approximated by a reduced model, the bi-stability was explored through the use of Dynamic Mode Decomposition (DMD). It was determined that the phenomenon is driven by primarily by the first two modal pairs which encapsulate 24.5\%  of the modal influence are associated either with the primary pressure structure, or its initial redistribution. With higher modes leading to the its fragmentation or in its transition vortex shedding. 

The time development of these modes further elucidated on the effect. For the first modal pair, as the cycle progressed, pressure propagation between the shock and separation bubble caused interference, resulting in shock detachment from the surface. This mechanism was responsible for both the detachment and subsequent reattachment of the jet. The second modal pair complemented the first by an additional feedback mechanism that drives the pressure redistribution.

These findings reveal that under the examined conditions, bi-stability fundamentally disrupts the circulation control mechanism. Yet, despite its complexity, the phenomenon is accurately captured using URANS and distilled even further into a ROM model. Remarkably, only six dynamic modes are sufficient to reconstruct the unsteady pressure field with 98\%, offering a powerful, efficient path toward its modeling